\documentclass[12pt]{article}

\pdfoutput=1 
\usepackage{graphicx}
\usepackage{xspace}
\usepackage{amsmath}
\usepackage{hyperref} 
\usepackage{natbib}
\usepackage{url}

\newcommand{\JL}{{JL}\xspace}
\newcommand{\NP}{{NP}\xspace}
\newcommand{\SM}{{SM}\xspace}
\newcommand{\BF}{{\rm BF}\xspace}
\newcommand{\kl}{\ensuremath{\mathrm{K^0_L}}}
\newcommand{\ks}{\ensuremath{\mathrm{K^0_S}}}
\newcommand{\zzero}{\ensuremath{\mathrm{Z^0}}}
\newcommand{\mz}{\ensuremath{m_\mathrm{Z}}}
\newcommand{\zp}{\ensuremath{\mathrm{Z^\prime}}}
\newcommand{\pmin}{\ensuremath{p\mathrm{_{min}}}}
\newcommand{\Xbar}{{\overline{X}}}
\newcommand{\xbar}{{\overline{x}}}
\newcommand{\sigmatot}{\ensuremath{\sigma_{\mathrm{tot}}}\xspace}

\newcommand{\quarkb}{b}
\newcommand{\sbar}{\ensuremath{\overline{\mathrm{s}}}} 
\newcommand{\bs}{\ensuremath{\mathrm{B_s}}\xspace} 

\begin{document}

\title{\bf The Jeffreys-Lindley Paradox and \\
Discovery Criteria in High Energy Physics}
\author{Robert D. Cousins\thanks{cousins@physics.ucla.edu}
\\ Department of Physics and Astronomy\\ University of California,
Los Angeles, California 90095, USA}

\date{August 23, 2014}

\maketitle

\begin{abstract}
  The Jeffreys--Lindley paradox displays how the use of a $p$-value
  (or number of standard deviations $z$) in a frequentist hypothesis
  test can lead to an inference that is radically different from that
  of a Bayesian hypothesis test in the form advocated by Harold
  Jeffreys in the 1930s and common today.  The setting is the test of
  a well-specified null hypothesis (such as the Standard Model of
  elementary particle physics, possibly with ``nuisance parameters'')
  versus a composite alternative (such as the Standard Model plus a
  new force of nature of unknown strength).  The $p$-value, as well
  as the ratio of the likelihood under the null hypothesis to the
  maximized likelihood under the alternative, can strongly
  disfavor the null hypothesis, while the Bayesian posterior
  probability for the null hypothesis can be arbitrarily large.  The
  academic statistics literature contains many impassioned comments on
  this paradox, yet there is no consensus either on its relevance to
  scientific communication or on its correct resolution.  The paradox
  is quite relevant to frontier research in high energy physics. This
  paper is an attempt to explain the situation to both physicists and
  statisticians, in the hope that further progress can be made.
\end{abstract}

\clearpage
\tableofcontents
\clearpage

\section{Introduction}
\label{intro}
On July 4, 2012, the leaders of two huge collaborations (CMS and
ATLAS) presented their results at a joint seminar at the CERN
laboratory, located on the French--Swiss border outside Geneva. Each
described the observation of a ``new boson'' (a type of particle),
suspected to be the long-sought Higgs boson \citep*{july4}.  The
statistical significances of the results were expressed in terms of
``$\sigma$'': carefully calculated $p$-values (not assuming normality)
were mapped onto the equivalent number of standard deviations in a
one-tailed test of the mean of a normal (i.e., Gaussian) distribution.
ATLAS observed 5$\sigma$ significance by combining the two most
powerful detection modes (different kinds of particles into which the
boson decayed) in 2012 data with full results from earlier data. With
independent data from a different apparatus, and only partially
correlated analysis assumptions, CMS observed 5$\sigma$ significance
in a similar combination, and when combining with some other modes as
CMS had planned for that data set, 4.9$\sigma$.

With ATLAS and CMS also measuring similar values for the rates of
production of the detected particles, the new boson was immediately
interpreted as the most anticipated and publicized discovery in high
energy physics (HEP) since the Web was born (also at CERN).
Journalists went scurrying for explanations of the meaning of
``$\sigma$'', and why ``high energy physicists require 5$\sigma$ for a
discovery''.  Meanwhile, some who knew about Bayesian hypothesis
testing asked why high energy physicists were using frequentist
$p$-values rather than calculating the posterior belief in the
hypotheses.

In this paper, I describe some of the traditions for claiming
discovery in HEP, which have a decidedly frequentist flavor, drawing
in a pragmatic way on both Fisher's ideas and the Neyman--Pearson (NP)
approach, despite their disagreements over foundations of statistical
inference.  Of course, some HEP practitioners have been aware of the
criticisms of this approach, having enjoyed interactions with some of
the influential Bayesian statisticians (both subjective and objective
in flavor) who attended HEP workshops on statistics.  These issues
lead directly to a famous ``paradox'', as \citet{lindley1957} called
it, when testing the hypothesis of a specific value $\theta_0$ of a
parameter against a continuous set of alternatives $\theta$.  The
different scaling of $p$-values and Bayes factors with sample size,
described by Jeffreys and emphasized by Lindley, can lead the
frequentist and the Bayesian to inconsistent strengths of inferences
that in some cases can even reverse the apparent inferences.

However, as described below, it is an understatement to say that the
community of Bayesian statisticians has not reached full agreement on
what should replace $p$-values in scientific communication. For
example, two of the most prominent voices of ``objective'' Bayesianism
(J. Berger and J. Bernardo) advocate fundamentally different
approaches to hypothesis testing for scientific communication.
Furthermore, views in the Bayesian literature regarding the validity
of models (in the social sciences for example) are strikingly
different than those common in HEP.

This paper describes today's rather unsatisfactory situation.
Progress in HEP meanwhile continues, but it would be
potentially quite useful if more statisticians become aware of the
special circumstances in HEP, and reflect on what the
Jeffreys--Lindley (JL) paradox means to HEP, and vice versa.

In ``high energy physics'', also known as ``elementary particle
physics'', the objects of study are the smallest building blocks of
matter and the forces among them.  (For one perspective, see
\citet{wilczek2004}.)  The experimental techniques often make use of
the highest-energy accelerated beams attainable. But due to the magic
of quantum mechanics, it is possible to probe much higher energy
scales through precise measurements of certain particle decays at
lower energy; and since the early universe was hotter than our most
energetic beams, and still has powerful cosmic accelerators and
extreme conditions, astronomical observations are another crucial
source of information on ``high energy physics''.
Historically, many discoveries in HEP have been in the category known
to statisticians as ``the interocular traumatic test; you know what
the data mean when the conclusion hits you between the eyes.''
\citep[p.~217, citing J. Berkson]{edwards1963}.  In other cases,
evidence accumulated slowly, and it was considered essential to
quantify evidence in a fashion that relates directly to the subject of
this review.

A wide range of views on the \JL paradox can be found in reviews with
commentary by many distinguished statisticians, in particular those of
\citet{shafer1982}, \citet{bergersellke1987},
\citet{bergerdelampady1987}, and \citet*{robert2009}.  The review of
Bayes factors by \citet{kassraftery1995} and the earlier book by
economist \citet{leamer1978} also offer interesting insights.  Some of
these authors view statistical issues in their typical data analyses
rather differently than do physicists in HEP; perhaps the greatest
contrast is that physicists {\em do} often have non-negligible belief
that their null hypotheses are valid to a precision much greater than
our measurement capability.  Regarding the search by ATLAS and CMS
that led to the discovery of ``a Higgs boson'', statistician
\citet{vandyk2014} has prepared an informative summary of the
statistical procedures that were used.

In Sections~\ref{original}--\ref{threescales}, I review the paradox,
discuss the concept of the point null hypothesis, and observe that the
paradox arises if there are three different scales in $\theta$ having
a hierarchy that is common in HEP.  In Section~\ref{wrong}, I address
the notions common among statisticians that ``all models are wrong'',
and that scientists tend to be biased against the null hypothesis, so
that the paradox is irrelevant.  I also describe the likelihood-ratio
commonly used in HEP as the test statistic. In Section~\ref{scaletau},
I discuss the difficult issue of choosing the prior for $\theta$, and
in particular the scale $\tau$ of those values of $\theta$ for which
there is non-negligible prior belief.  Section~\ref{bernardo} briefly
describes the completely different approach to hypothesis testing
advocated by Bernardo, which stands apart from the bulk of the
Bayesian literature. In Section~\ref{effectsize}, I discuss how
measured values and confidence intervals, for quantities such as
production and decay rates, augment the quoted $p$-value, and how
small but precisely measured effects can provide a window into {\em
  very} high energy physics.  Section~\ref{npalpha} discusses the
choice of Type I error $\alpha$ (probability of rejecting $H_0$ when
it is true) when adopting the approach of \NP hypothesis testing, with
some comments on the ``5$\sigma$ myth'' of HEP.  Finally, in
Section~\ref{pvaluesummary}, I discuss the seemingly universal
agreement that a single $p$-value is (at best) a woefully incomplete
summary of the data, and how confidence intervals at various
confidence levels help readers assess the experimental results.  I
summarize and conclude in Section~\ref{conclusion}.

As it is useful to use precisely defined terms, we must be aware that
statisticians and physicists (and psychologists, etc.) have different
naming conventions. For example, a physicist says ``measured value'',
while a statistician says ``point estimate'' (and while a psychologist
says ``effect size in original units'').  This paper uses primarily
the language of statisticians, unless otherwise stated. Thus
``estimation'' does not mean ``guessing'', but rather the calculation
of ``point estimates'' and ``interval estimates''.  The latter refers
to frequentist confidence intervals or their analogs in other
paradigms, known to physicists as ``uncertainties on the measured
values''.  In this paper, ``error'' is generally used in the precisely
defined sense of Type I and Type II errors of Neyman-Pearson theory
(Section~\ref{npalpha}), unless obvious from context.  Other terms are
defined in context below.  Citations are provided for the benefit of
readers who may not be aware that certain terms (such as ``loss'')
have specific technical meanings in the statistics literature.
``Effect size'' is commonly used in the psychology literature, with at
least two meanings.  The first meaning, described by the field's
publication manual \citep[p.~34]{apa2010} as ``most often easily
understood'', is simply the measured value of a quantity in the
original (often dimensionful) units.  Alternatively, a
``standardized'' dimensionless effect size is obtained by dividing by
a scale such as a standard deviation.  In this paper, the term always
refers to the former definition (original units), corresponding to the
physicist's usual measured value of a parameter or physical quantity.
Finally, the word ``model'' in statistics literature usually refers to
a probabilistic equation that describes the assumed data-generating
mechanisms (Poisson, binomial, etc.), often with adjustable
parameters.  The use of ``model'' for a ``law of nature'' is discussed
below.

\section{The original ``paradox'' of Lindley, as corrected by Bartlett}
\label{original}
\begin{figure}
\centering
\includegraphics[width=\textwidth]{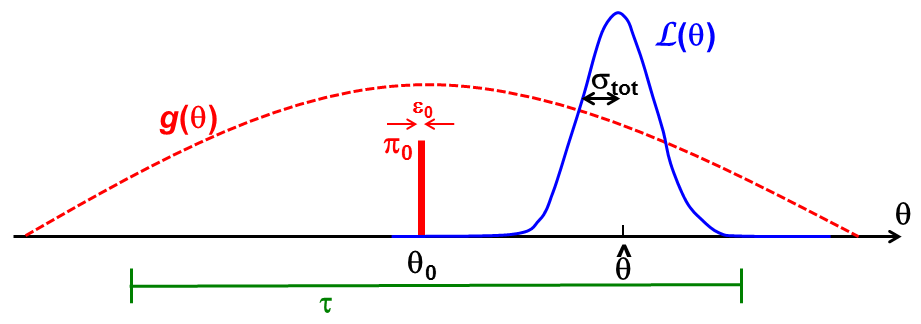}
\caption{Illustration of quantities used to define the 
JL paradox.  The unknown parameter is $\theta$, with likelihood
function ${\cal L}(\theta)$ resulting from a measurement with
uncertainty $\sigmatot$.  The point MLE is $\hat\theta$, which in the
sketch is about $5\sigmatot$ away from the null hypothesis, the
``point null'' $\theta_0$.  The point null hypothesis has prior
probability $\pi_0$, which can be spread out over a small interval of
width $\epsilon_0$ without materially affecting the paradox.  The
width of the prior pdf $g(\theta)$ under $H_1$ has scale $\tau$.  The
scales have the hierarchy $\epsilon_0 \ll \sigmatot \ll \tau$.
\label{fig:JL}
}
\end{figure}

\citet{lindley1957}, with a crucial correction by
\citet{bartlett1957}, lays out the paradox in a form that is useful as
our starting point.  This exposition also draws on Section 5.0 of
\citet{jeffreys1961} and on \citet*{bergerdelampady1987}.  It mostly
follows the notation of the latter, with the convention of upper case
for the random variable and lower case for observed values.
Figure~\ref{fig:JL} serves to illustrate various quantities defined
below.

Suppose $X$ having density $f(x|\theta)$ is sampled, where $\theta$ is
an unknown element of the parameter space $\Theta$. It is desired to
test $H_0$: $\theta=\theta_0$ versus $H_1$: $\theta\ne\theta_0$.
Following the Bayesian approach to hypothesis testing pioneered by
Jeffreys (also referred to as Bayesian model selection), we assign
prior probabilities $\pi_0$ and $\pi_1 = 1 - \pi_0$ to the respective
hypotheses.  Conditional on $H_1$ being true, one also has a
continuous prior probability density $g(\theta)$ for the unknown
parameter.

As discussed in the following sections, formulating the problem in
this manner leads to a conceptual issue, since in the continuous
parameter space $\Theta$, a single point $\theta_0$ (set of measure
zero) has non-zero probability associated with it.  This is impossible
with a usual probability density, for which the probability assigned
to an interval tends to zero as the width of the interval tends to
zero.  Assignment of non-zero probability $\pi_0$ to a single point
$\theta_0$ is familiar to physicists by using the Dirac
$\delta$-function (times $\pi_0$) at $\theta_0$, while statisticians
often refer to placing ``probability mass'' at $\theta_0$, or to using
``counting measure'' for $\theta_0$ (in distinction to ``Lebesgue
measure'' for the usual density $g$ for $\theta\ne\theta_0$).  The
null hypothesis corresponding to the single point $\theta_0$ is also
commonly referred to as a ``point null'' hypothesis, or as a ``sharp
hypothesis''.  As discussed below, just as a $\delta$-function can be
viewed as useful approximation to a highly peaked function, for
hypotheses in HEP it is often the case that the point null hypothesis
is a useful approximation to a prior that is sufficiently concentrated
around $\theta_0$.

If the density $f(x|\theta)$ under $H_1$ is normal with mean $\theta$
and known variance $\sigma^2$, then for a random sample $\{x_1, x_2,
\dots x_n\}$, the sample mean is normal with variance $\sigma^2/n$,
i.e., $\Xbar$ has density $N(\theta, \sigma^2/n)$.  For conciseness
(and eventually to make the point that ``$n$'' can be obscure), let
\begin{equation}
\label{eqn:sigmatot}
\sigmatot \equiv \sigma/\sqrt{n}.  
\end{equation}
The likelihood is then
\begin{equation}
\label{like}
{\cal L}(\theta) = \frac{1}{\sqrt{2\pi}\sigmatot} 
  \exp\left\{-(\xbar - \theta)^2/2\sigmatot^2\right\},
\end{equation}
with maximum likelihood estimate (MLE) $\hat\theta = \xbar$.  By
Bayes's Theorem, the posterior probabilities of the hypotheses, given
$\hat\theta$, are:
\begin{equation}
P(H_0|\hat\theta)  = \frac{1}{A}\,\pi_0\,{\cal L}(\theta_0)
                   = \frac{1}{A}\,\pi_0\,\frac{1}{\sqrt{2\pi}\sigmatot} 
  \exp\left\{-(\hat\theta - \theta_0)^2/2\sigmatot^2\right\}
\end{equation}
and
\begin{equation}
\label{pH1}
P(H_1|\hat\theta)  
 = \frac{1}{A}\,\pi_1\,\int g(\theta) {\cal L}(\theta)  d\theta
 = \frac{1}{A}\,\pi_1\,\int g(\theta) \frac{1}{\sqrt{2\pi}\sigmatot} 
 \exp\left\{-(\hat\theta - \theta)^2/2\sigmatot^2\right\} d\theta.
 \end{equation}
Here $A$ is a normalization constant to make the sum of the two
probabilities equal unity, and the integral is over the support of
the prior $g(\theta)$.

There will typically be a scale $\tau$ that indicates the range of
values of $\theta$ over which $g(\theta)$ is relatively large.  One
considers the case
\begin{equation}
\label{sigmatau}
\sigmatot \ll \tau,
\end{equation}
so that $g(\theta)$ varies slowly where the rest of the integrand is
non-negligible, and therefore the integral approximately equals
$g(\hat\theta)$, so that
\begin{equation}
P(H_1|\hat\theta)  
 \approx \frac{1}{A}\,\pi_1\,g(\hat\theta)
\end{equation}
Then the ratio of posterior odds to prior odds for $H_0$, i.e., the
{\em Bayes factor} (\BF), is independent of $A$ and $\pi_0$, and given
by
\begin{align}
\BF &\equiv \frac{P(H_0|\hat\theta)}{P(H_1|\hat\theta)} \bigg / \frac{\pi_0}{\pi_1} \approx
  \frac{1}{\sqrt{2\pi}\sigmatot g(\hat\theta)} 
  \exp\left\{-(\hat\theta - \theta_0)^2/2\sigmatot^2\right\} \nonumber \\
  &=   \frac{1}{\sqrt{2\pi}\sigmatot g(\hat\theta)}  \exp(-z^2/2),
\end{align}
where 
\begin{equation}
z = (\hat\theta - \theta_0)/\sigmatot = \sqrt{n}(\hat\theta - \theta_0)/\sigma
\end{equation}
is the usual statistic providing the departure from the null
hypothesis in units of $\sigmatot$.  Some authors (e.g.,
\citet{kassraftery1995}) use the notation $B_{01}$ for this Bayes
factor, to make clear which hypotheses are used in the ratio; as this
paper always uses the same ratio, the subscripts are suppressed. Then
the $p$-value for the two-tailed test is $p=2(1 - \Phi(z))$, where
$\Phi$ is the standard normal cumulative distribution function. (As
discussed in Section~\ref{heptest}, in HEP often $\theta$ is
physically non-negative, and hence a one-tailed test is used, i.e.,
$p=1 - \Phi(z)$.)

\citet[p.~248]{jeffreys1961} notes that $g(\hat\theta)$ is independent
of $n$ and $\sigmatot$ goes as $1/\sqrt{n}$, and therefore a given
cutoff value of \BF does {\em not} correspond to a fixed value of $z$.
This discrepancy in the sample-size scaling of $z$ and $p$-values
compared to that of Bayes factors (already noted for a constant $g$ on
p.~194 in his first edition of 1939) is at the core of the \JL
paradox, even if one does not take values of $n$ so extreme as to make
$P(H_0|\hat\theta)>P(H_1|\hat\theta)$.

\citet[Appendix B, p.~435]{jeffreys1961} curiously downplays the
discrepancy at the end of a sentence that summarizes his objections to
testing based on $p$-values (almost verbatim with p.~360 of his 1939
edition): ``In spite of the difference in principle between my tests
and those based on [$p$-values], and the omission of the latter to
give the increase in the critical values for large $n$, dictated
essentially by the fact that in testing a small departure found from a
large number of observations we are selecting a value out of a long
range and should allow for selection, it appears that there is not
much difference in the practical recommendations.''  He does say, ``At
large numbers of observations there is a difference'', but he suggests
that this will be rare and that the test might not be properly
formulated: ``internal correlation should be suspected and tested''.

In contrast, \citet{lindley1957} emphasized how large the discrepancy
could be, using the example where $g(\theta)$ is taken to be constant
over an interval that contains both $\hat\theta$ and the range of
$\theta$ in which the integrand is non-negligible.  For any
arbitrarily small $p$-value (arbitrarily large $z$) that is
traditionally interpreted as evidence {\em against the null
  hypothesis}, there will always exist $n$ for which the \BF can be
{\em arbitrarily large in favor of the null hypothesis}.

\citet{bartlett1957} quickly noted that Lindley had neglected the
length of the interval over which $g(\theta)$ is constant, which
should appear in the numerator of the \BF, and which makes the
posterior probability of $H_0$ ``much more arbitrary''.  More
generally, the normalization of $g$ always has a scale $\tau$ that
characterizes the extent in $\theta$ for which $g$ is non-negligible,
which implies that $g(\hat\theta) \propto 1/\tau $. Thus, there is a
factor of $\tau$ in the numerator of \BF.  For example,
\citet{bergerdelampady1987} and others consider $g(\theta)$ having
density $N(\theta_0, \tau^2)$, which, in the limit of
Eqn.~\ref{sigmatau}, leads to
\begin{equation}
\BF =  \frac{\tau}{\sigmatot}   \exp(-z^2/2).
\end{equation}
There is the same proportionality in the Lindley/Bartlett example if
the length of their interval is $\tau$. The crucial point is the
generic scaling,
\begin{equation}
\label{scaling}
\BF \propto  \frac{\tau}{\sigmatot}  \exp(-z^2/2).
\end{equation}
Of course, the value of the proportionality constant depends on the
form of $g$ and specifically on $g(\hat\theta)$.

Meanwhile, from Eqn.~\ref{like}, the ratio $\lambda$ of the likelihood
of $\theta_0$ under $H_0$ and the maximum likelihood under $H_1$ is
\begin{align}
\lambda &= {\cal L}(\theta_0) / {\cal L}(\hat\theta) \\
        &=\exp\left\{(\hat\theta - \theta_0  )^2/2\sigmatot^2\right\} \bigg /
 \exp\left\{(\hat\theta - \hat\theta)^2/2\sigmatot^2\right\} \\
 &=  \exp(-z^2/2)  \\
 &\propto \left( \frac{\sigmatot}{\tau} \right) \BF. 
\label{ockham}
\end{align}
Thus, {\em unlike} the case of simple-vs-simple hypotheses discussed
below in Section~\ref{simple}, this maximum likelihood ratio takes the
side of the $p$-value in disfavoring the null hypothesis for large
$z$, independent of $\sigmatot/\tau$, and thus independent of sample
size $n$.  This difference between maximizing ${\cal L}(\theta)$ under
$H_1$, and averaging it under $H_1$ weighted by the prior $g(\theta)$,
can be dramatic.

The factor $\sigmatot/\tau$ (arising from the average of $\cal L$
weighted by $g$ in Eqn.~\ref{pH1}) is often called the ``Ockham
factor'' that provides a desirable ``Ockham's razor'' effect
\citep[Chapter 20]{jaynes2003} by penalizing $H_1$ for imprecise
specification of $\theta$.  But the fact that (even asymptotically) BF
depends directly on the scale $\tau$ of the prior $g(\theta)$ (and
more precisely on $g(\hat\theta)$) can come as a surprise to those
deeply steeped in Bayesian point and interval estimation, where
typically the dependence on all priors diminishes asymptotically.  The
surprise is perhaps enhanced since the BF is often introduced as the
factor by which prior odds (even if subjective) are modified in light
of the observed data, giving the initial impression that the
subjective part is factorized out from the \BF.

The likelihood ratio $\lambda = \exp(-z^2/2)$ takes on the numerical
values 0.61, 0.14, 0.011, 0.00034, and 3.7E-06, as $z$ is equal to 1,
2, 3, 4, and 5, respectively.  Thus, in order for the Ockham factor to
reverse the preferences of the hypotheses in the BF compared to the
maximum likelihood ratio $\lambda$, the Ockham factor must be smaller
than these numbers in the respective cases.  Some examples of
$\sigmatot$ and $\tau$ in HEP that can do this (at least up to $z=4$)
are in Section~\ref{hepex}.  As discussed below, even when not in the
extreme case where the Ockham factor reverses the preference of the
hypotheses, its effect deserves scrutiny.

From the derivation, the origin of the Ockham factor (and hence
sample-size dependence) does not depend on the chosen value of
$\pi_0$, and thus not on the commonly suggested choice of $\pi_0 =
1/2$.  The scaling in Eqn.~\ref{scaling} follows from assigning {\em
  any} non-zero probability to the single point $\theta=\theta_0$, as
described above using the Dirac $\delta$-function, or ``probability
mass''.

The situation clearly invited further studies, and various authors,
beginning with \citet{edwards1963}, have explored the impact of
changing $g(\theta)$, making numerical comparisons of $p$-values to
Bayes factors in contexts such as testing a point null hypothesis for
a binomial parameter.  Generally they have given examples in which the
$p$-value is always numerically smaller than the BF, even when the
prior for $\theta$ ``gives the utmost generosity to the alternative
hypothesis''.

\subsection{Is there really a ``paradox''?}

A trivial ``resolution'' of \JL paradox is to point out that there is
no reason to expect the numerical results of frequentist and Bayesian
hypothesis testing to agree, as they calculate different quantities.
Still, it is unnerving to many that ``hypothesis tests'' that are both
communicating scientific results for the same data can have such a
large discrepancy. So is it a {\em paradox}?

I prefer to use the word ``paradox'' with the meaning I recall from
school, ``a statement that is seemingly contradictory or opposed to
common sense and yet is perhaps true'' \citep[definition
2a]{webster7}.  This is the meaning of the word, for example, in the
celebrated ``paradoxes'' of Special Relativity, such as the Twin
Paradox and the Pole-in-Barn Paradox.  The ``resolution'' of a paradox
is then a careful explanation of why it is {\em not} a contradiction.
I therefore do {\em not} use the word paradox as a synonym for
contradiction---that takes a word with (I think) a very useful meaning
and wastes it on a redundant meaning of another word.  It can however
be confusing that what is deemed paradoxical depends on the personal
perspective of what is ``seemingly'' contradictory. If someone says,
``What Lindley called a paradox is not a paradox'', then typically
they either define paradox as a synonym for contradiction, or it was
always so obvious to them that the paradox is not a contradiction that
they think it is not paradoxical.  (It could also be that there {\em
  is} a contradiction that cannot be resolved, but I have not seen
that used as an argument for why it is not a paradox.)  Although it
may still be questionable as to whether there is a resolution
satisfactory to everyone, for now I think that the word paradox is
quite apt.  As the deep issue is the scaling of the BF with sample
size (for fixed $p$-value) as pointed out by Jeffreys already in 1939,
I follow some others in calling it the Jeffreys--Lindley (JL) paradox.

Other ambiguities in discussions regarding the JL paradox include
whether the focus is on the posterior odds of $H_0$ (which includes
the prior odds) or on the BF (which does not).  In addition, while one
often introduces the paradox by noting the extreme cases where the
$p$-value and the BF seem to imply opposite inferences, one should
also emphasize the less dramatic (but still disturbing) cases where
the Ockham factor plays a large (and potentially) arbitrary role, even
if the BF favors $H_1$.  In the latter cases, it can be claimed that
the $p$-value overstates the evidence against $H_0$.  In this paper I
focus on the BF, following some others, e.g.  \citet[who somewhat
confusingly denote it by $L$, p.~218]{edwards1963} and \citet[p.
102]{bernardo1999bayes}.  I also take a rather inclusive view of the
paradox, as the issue of differences in sample size scaling is always
present, even if not taken to the extreme limit where the Ockham
factor overwhelms the BF, and even reverses arbitrarily small prior
probability for $H_0$.

\subsection{The JL paradox is \texorpdfstring{\textit{not}}{not} 
about testing  \texorpdfstring{simple $H_0$ vs simple $H_1$}
{simple H0 vs simple H1}}
\label{simple}
Testing simple $H_0$: $\theta=\theta_0$ vs {\em simple} $H_1$: $\theta
= \theta_1$ provides another interesting contrast between Bayesian and
frequentist hypothesis testing, but this is {\em not} an example of
the \JL paradox.  The Bayes factor and the likelihood ratio are the
{\em same} (in the absence of nuisance parameters), and therefore in
agreement as to which hypothesis the data favor.  This is in contrast
to the high-$n$ limit of the \JL paradox,

In the situation of the \JL paradox, there is a value of $\theta$
under $H_1$ that is {\em equal} to the MLE $\hat\theta$, and which
consequently has a likelihood no lower than that of $\theta_0$.  The
extent to which $\hat\theta$ is not favored by the prior is encoded in
the Ockham factor of Eqn.~\ref{ockham}, which means that the BF and
the likelihood ratio $\lambda$ can disagree on both the magnitude and
even the direction of the evidence.

Simple-vs-simple hypothesis tests are far less common in HEP than
simple-vs-composite tests, but have arisen as the CERN experiments
have been attempting to infer properties of the new boson, such as the
quantum numbers that characterize its spin and parity. Again supposing
$X$ having density $f(x|\theta)$ is sampled, now one can form {\em
  two} well-defined $p$-values, namely $p_0$ indicating departures
from $H_0$ in the direction of $H_1$, and $p_1$ indicating departures
from $H_1$ in the direction of $H_0$.  A physicist will examine both
$p$-values in making an inference.

\citet[p.~108]{thompson2007} argues that the set of the {\em two}
$p$-values is ``the evidence'', and many in HEP may agree.  Certainly
neglecting one of the $p$-values can be dangerous.  For example, if
$\theta_0 < \hat\theta < \theta_1$, and $\sigmatot \ll \theta_1 -
\theta_0$, then it is conceivable that $H_0$ is rejected at 5$\sigma$,
while if $H_1$ were the null hypothesis, it would be rejected at
7$\sigma$.  A physicist would be well aware of this circumstance and
hardly fall into the straw-man trap of implicitly accepting $H_1$ by
focusing only on $p_0$ and ``rejecting'' (only) $H_0$.  The natural
reaction would be to question both hypotheses; i.e., the
two-simple-hypothesis model would be questioned.  (In this context,
\citet[pp.~200-201]{senn2001} has further criticism and references
regarding the issue of sample-size dependence of $p$-values.)

\section{Do point null hypotheses make sense in principle, or in practice?}
\label{pointnull}

In the Bayesian literature, there are notably differing attitudes
expressed regarding the relevance of a point null hypothesis
$\theta=\theta_0$.  Starting with Jeffreys, the fact that Bayesian
hypothesis testing can treat a point null hypothesis in a special way
is considered by many proponents to be an advantage.  (As discussed in
Section~\ref{npalpha}, frequentist testing of a point null vs a
composite alternative is tied to interval estimation, a completely
different approach.)  The hypothesis test is often phrased in the
language of model selection: the ``smaller'' model $H_0$ is nested in
the ``larger'' model $H_1$.  From this point of view, it seems natural
to have one's prior probabilities $\pi_0$ and $\pi_1$ for the two
models.  However, as mentioned above, from the point of view of
putting a prior on the entire space $\Theta$ in the larger model, this
corresponds to a non-regular prior that has counting measure
($\delta$-function to physicists) on $\theta_0$ and Lebesgue measure
(usual probability density to physicists) on $\theta\ne\theta_0$.

As discussed by \citet{casellaberger1987}, some of the more disturbing
aspects of the \JL paradox are ameliorated (or even ``reconciled'') if
there is no point null, and the test is the so-called ``one-sided
test'', namely $H_0$: $\theta\le\theta_0$ vs $H_1$: $\theta >
\theta_0$.  Given the importance of the issue of probability assigned
to the point null, some of the opinions expressed in the statistics
literature are highlighted below, to contrast with the attitude in HEP
described in Section~\ref{wrong}.

\citet{lindley2009} lauds the ``triumph'' of Jeffreys's ``general
method of significance tests, putting a concentration of prior
probability on the null---no ignorance here---and evaluating the
posterior probability using what we now call Bayes factors.''  As a
strong advocate of the use of subjective priors that represent
personal belief, Lindley views the probability mass on the point null
as subjective. (In the same comment, Lindley criticizes Jeffrey's
``error'' of integrating over the sample space of unobserved data in
formulating his eponymous priors for use in point and interval
estimation.)

At the other end of the spectrum of Bayesian theorists,
\citet{bernardo2009} comments on \citet{robert2009}: ``Jeffreys
intends to obtain a posterior probability for a precise null
hypothesis, and, to do this, he is forced to use a mixed prior which
puts a lump of probability $p=Pr(H_0)$ on the null, say $H_0 \equiv
{\theta=\theta_0}$ and distributes the rest with a {\em proper} prior
$p(\theta)$ (he mostly chooses $p=1/2$).  This has a very upsetting
consequence, usually known as Lindley's paradox: for any fixed prior
probability $p$ independent of the sample size $n$, the procedure will
wrongly accept $H_0$ whenever the likelihood is concentrated around a
true parameter value which lies $O(n^{-1/2})$ from $H_0$.  I find it
difficult to accept a procedure which is {\em known} to produce the
wrong answer under specific, but not controllable, circumstances.''
When pressed by commenters, \cite{bernardo2011bayes} says that ``I am
sure that there are situations where the scientist is willing to use a
prior distribution highly concentrated at a particular region and
explore the consequences of this assumption\dots{}What I claim is
that, even in precise hypothesis testing situations, the scientist is
often interested in an analysis which does {\em not} assume this type
of sharp prior knowledge\dots .''  Bernardo goes on to advocate a
different approach (Section~\ref{bernardo}), which ``has the
nontrivial merit of being able to use for both estimation and
hypothesis testing problems a single, unified theory for the
derivation of objective `reference' priors.''

Some statisticians find point null hypotheses irrelevant to their own
work.  In the context of an unenthusiastic comment on the Bayesian
information criterion (BIC), \citet*{gelmanrubin1995} say ``More
generally, realistic prior distributions in social science do not have
a mass of probability at zero\dots .'' \citet{raftery1995r} disagrees,
saying that ``social scientists are prepared to act {\em as if} they
had prior distributions with point masses at zero\dots{}social
scientists often entertain the possibility that an effect is {\em
  small}\,''.

In the commentary of \citet{bernardo2011bayes}, C. Robert and J.
Rousseau say, ``{\em Down with point masses!} The requirement that one
uses a point mass as a prior when testing for point null hypotheses is
always an embarrassment and often a cause of misunderstanding in our
classrooms.  Rephrasing the decision to pick the simpler model as the
result of a larger advantage is thus much more likely to convince our
students. What matters in pointwise hypothesis testing is not whether
or not $\theta=\theta_0$ holds but what the consequences of a wrong
decision are.''

Some comments on the point null hypothesis are related to another
claim, that all models and all point nulls are at best approximations
that are wrong at some level.  I discuss this point in more detail in
Section~\ref{wrong}, but include a few quotes here.
\citet{edwards1963} say, ``\dots{}in typical applications, one of the
hypotheses---the null hypothesis---is known by all concerned to be
false from the outset,'' citing others including \citet{berkson1938}.
\citet{vardeman1987} claims, ``Competent scientists do not believe
their own models or theories, but rather treat them as convenient
fictions.  A small (or even 0) prior probability that the current
theory is true is not just a device to make posterior probabilities as
small as $p$ values, it is the way good scientists think!''

\citet*{casellaberger1987c} object specifically to Jeffreys's use of
$\pi_0=\pi_1=1/2$, used in modern papers as well: ``Most researchers
would not put 50\% prior probability on $H_0$.  The purpose of an
experiment is often to disprove $H_0$ and researchers are not
performing experiments that they believe, {\em a priori}, will fail
half the time!''  \citet{kadane1987} expresses a similar sentiment:
``For the last 15 years or so I have been looking seriously for
special cases in which I might have some serious belief in a null
hypothesis. I have found only one [testing astrologer]\dots I do not
expect to test a precise hypothesis as a serious statistical
calculation.''

As discussed below, such statisticians have evidently not been
socializing with many HEP physicists. In fact, in the literature I
consulted, I encountered very few statisticians who granted, as did
\citet{zellner2009}, that physical laws such as $E=mc^2$ are point
hypotheses, and ``Many other examples of sharp or precise hypotheses
can be given and it is incorrect to exclude such hypotheses {\em a
  priori} or term them `unrealistic'\dots .''

Condensed matter physicist and Nobel Laureate Philip
\citet{anderson1992} argued for Jeffreys-style hypothesis testing with
respect to a claim for evidence for a fifth force of nature.  ``Let us
take the `fifth force'.  If we assume from the outset that there {\em
  is} a fifth force, and we need only measure its magnitude, we are
assigning the bin with zero range and zero magnitude an infinitesimal
probability to begin with.  Actually, we should be assigning this bin,
which is the null hypothesis we want to test, some {\em finite a
  priori} probability---like 1/2---and sharing out the remaining 1/2
among all the other strengths and ranges.''

Already in \citet[p.~235]{edwards1963} there was a key point related
to the situation in HEP: ``Bayesians\dots{}must remember that the null
hypothesis is a hazily defined small region rather than a point.''
They also emphasized the subjective nature of singling out a point
null hypothesis: ``At least for Bayesian statisticians, however, no
procedure for testing a sharp null hypothesis is likely to be
appropriate unless the null hypothesis deserves special initial
credence.''

That the ``point'' null can really be a ``hazily defined small
region'' is clear from the derivation in Section~\ref{original}.  The
general scaling conclusion of Eqn.~\ref{scaling} remains valid if
``hazily defined small region'' means that the region of $\theta$
included in $H_0$ has a scale $\epsilon_0$ such that $\epsilon_0 \ll
\sigmatot$.  To a physicist, this just means that computing integrals
using a $\delta$-function is a good approximation to integrating over
a finite region in $\theta$.  (Some authors, such as
\citet{bergerdelampady1987} have explored quantitatively the
approximation induced in the \BF by non-zero $\epsilon_0$.)

\section{\texorpdfstring{\boldmath Three scales for $\theta$ yield a paradox}
              {Three scales for theta yield a paradox}}
\label{threescales}
From the preceding sections, we can conclude that for the \JL paradox
to arise, it is sufficient that there exist
{\em three scales} in the parameter space $\Theta$, namely:
\begin{enumerate}
\item $\epsilon_0$, the scale under $H_0$;
\item $\sigmatot$, the scale for the total measurement uncertainty; and
\item $\tau$, the scale under $H_1$; 
\end{enumerate}
and that they have the hierarchy
\begin{equation}
\label{eqn:scales}
\epsilon_0 \ll \sigmatot \ll \tau.
\end{equation}
This situation is common in frontier experiments in HEP, where, as
discussed in Section~\ref{hepex}, {\em the three scales are often
  largely independent}.  We even have cases where $\epsilon_0=0$,
i.e., most of the subjective prior probability is on $\theta=0$.  This
is the case if $\theta$ is the mass of the photon.

As noted for example by \citet{shafer1982}, the source of the
precision of $\sigmatot$ does not matter as long as condition in
Eqn.~\ref{eqn:scales} is satisfied.  The statistics literature tends
to focus on the case where $\sigmatot$ arises from a sample size $n$
via Eqn.~\ref{eqn:sigmatot}.  This invites the question as to whether
$n$ can really be arbitrarily large in order to make $\sigmatot$
arbitrarily small.  In my view the existence of a regime where the \BF
goes as $\tau/\sigmatot$ for fixed $z$ (as in Eqn.~\ref{scaling}) is
the fundamental characteristic that can lead to the \JL paradox, even
if this regime does not extend to $\sigmatot\rightarrow0$.  As I
discuss in Section~\ref{hepex}, such regimes are present in HEP
analyses, and there is not always a well-defined $n$ underlying
$\sigmatot$, a point I return to in Sections~\ref{heptest}
and~\ref{scaletau} below in discussing $\tau$.  But we first consider
the model itself.

\section{HEP and belief in the null hypothesis}
\label{wrong} 

At the heart of the measurement models in HEP are well-established
equations that are commonly known as ``laws of nature''.  By some
historical quirks, the current ``laws'' of elementary particle
physics, which have survived several decades of intense scrutiny with
only a few well-specified modifications, are collectively called a
``model'', namely the Standard Model (SM).  In this review, I refer to
the equations of such ``laws'', or alternatives considered as
potential replacements for them, as ``core physics models''.  The
currently accepted core physics models have parameters, such as masses
of the quarks and leptons, which with few exceptions have all been
measured reasonably precisely (even if requiring care to define).

Multiple complications arise in going from the core physics model to
the full measurement model that describes the probability densities
for observations such as the momentum spectra of particles emerging
from proton-proton collisions.  Theoretical calculations based on the
core physics model can be quite complex, requiring, for example,
approximations due to truncation of power series, incomplete
understanding of the internal structure of colliding protons, and
insufficient understanding of the manner in which quarks emerging from
the collision recombine into sprays of particles (``jets'') that can
be detected.  The results of such calculations, with their attendant
uncertainties, must then be propagated through simulations of the
response of detectors that are parametrized using many calibration
constants, adjustments for inefficient detection, misidentification of
particles, etc.  Much of the work in data analysis in HEP involves
subsidiary analyses to measure and calibrate detector responses, to
check the validity of theoretical predictions to describe data
(especially where no departures are expected), and to confirm the
accuracy of many aspects of the simulations.

The aphorism ``all models are wrong'' \citep{box1976} can certainly
apply to the detector simulation, where common assumptions of normal
or log-normal parametrizations are, at best, only good approximations.
But the pure core physics models still exist as testable hypotheses
that may be regarded as point null hypotheses.  Alternatives to the SM
are more generalized models in which the SM is nested.  It is
certainly worth trying to understand if some physical parameter in the
alternative core physics model is zero (corresponding to the SM), even
if it is necessary to do so through the smoke of imperfect detector
descriptions with many uninteresting and imperfectly known nuisance
parameters.  Indeed much of what distinguishes the capabilities of
experimenters is how well they can do precisely that by determining
the detector response through careful calibration and cross-checks.
This distinction is over-looked in the contention
\citep[p.~320]{bergerdelampady1987} that a point null hypothesis in a
core physics model cannot be precisely tested if the rest of the
measurement model is not specified perfectly.

There is a deeper point to be made about core physics models
concerning the difference between a model being a good
``approximation'' in the ordinary sense of the word, and the concept
of a mathematical limit.  The equations of Newtonian physics have been
superseded by those of special and general relativity, but the earlier
equations are not just approximations that did a good job in
predicting (most) planetary orbits; they are the correct {\em
  mathematical limits} in a precise sense.  The kinematic expressions
for momentum, kinetic energy, etc., are the limits of the special
relativity equations in the limit as the speed goes to zero.  That is,
if you specify a maximum tolerance for error due to the approximation
of Newtonian mechanics, then there exists a speed below which it will
always be correct within that tolerance.  Similarly, Newton's
universal law of gravity is the correct mathematical limit of General
Relativity in the limit of small gravitational fields and low speeds
(conditions that were famously not satisfied to observational
precision for the orbit of the planet Mercury).

This limiting behavior can often be viewed through an appropriate
power series.  For example, we can expand the expression for kinetic
energy $T$ from special relativity, $T = \sqrt{p^2+m^2} - m$, in
powers of $p^2/m^2$ in the non-relativistic limit where momentum $p$
is much smaller than the mass $m$.  The Newtonian expression,
$T=p^2/2m$, is the first term in the series, followed by the lowest
order relativistic correction term of $p^4/8m^3$. (I use the usual HEP
units in which the speed of light $c$ is 1 and dimensionless; to use
other units, substitute $pc$ for $p$, and $mc^2$ for $m$.)

An analogous, deeper concept arises in the context of effective field
theories.  An effective field theory in a sense consists of the
correct first term(s) in a power series of inverse powers of some
scale that is much higher than the applicable scale of the effective
theory \citep{georgi1993}.  When a theory is expressed as an infinite
series, a key issue is whether there is a {\em finite} number of
coefficients to be determined experimentally, from which all other
coefficients can be (at least in principle) calculated, with no
unphysical answers (in particular infinity) appearing for measurable
quantities.  Theories having this property are called {\em
  renormalizable}, and are naturally greatly favored over theories
that give infinities for measurable quantities or that require in
effect an infinite number of adjustable parameters.  It was a major
milestone in HEP theory when it was shown that the SM (including its
Higgs boson) is in a class of renormalizable theories
\citep{thooft1999}; removing the Higgs boson destroys this property.

In the last three or four decades, thousands of measurements have
tested the consistency of the predictions of the SM, many with
remarkable precision, including of course measurements at the LHC.
Nonetheless, the SM is widely believed to be incomplete, as it leaves
unanswered some obvious questions (such as why there are three
generations of quarks and leptons, and why their masses have the
values they do).  If the goal of a unified theory of forces is to
succeed, the current mathematical formulation will become embedded
into a larger mathematical structure, such that more forces and quanta
will have to be added.  Indeed much of the current theoretical and
experimental research program is aimed at uncovering these extensions,
while a significant effort is also spent on understanding further the
consequences of the known relationships.  Nevertheless, whatever new
physics is added, we also expect that the SM will remain a correct
mathematical limit, or a correct effective field theory, within a more
inclusive theory. It is in this sense of being the correct limit or
correct effective field theory that physicists believe that the SM is
``true'', both in its parts and in the collective whole.  (I am aware
that there are deep philosophical questions about reality, and that
this point of view can be considered ``naive'', but this is a point of
view that is common among high energy physicists.)

It may be that on deeper inspection the distinction between an
ordinary ``approximation'' and a mathematical limit will not be so
great, as even crude approximations might be considered as types
of limits.  Also, the usefulness of power series breaks down in
certain important ``non-perturbative'' regimes. Nonetheless, the
concepts of renormalizability, limits, and effective field theories
are helpful in clarifying what is meant by belief in core physics
models.  Comparing the approach of many physicists to that of
statisticians working in other fields, an important distinction
appears to be the absence of core ``laws'' in their models.  Under
such circumstances, one would naturally be averse to obsession about
exact values of model parameters when the uncertainty in the model
itself is already dominant.

\subsection{\texorpdfstring{Examples of three scales for $\theta$ in HEP experiments}
              {Examples of three scales for theta in HEP experiments}}
\label{hepex}

Many searches at the frontier of HEP have three scales with the
hierarchy in Eqn.~\ref{eqn:scales}.  An example is an experiment in
the 1980s that searched for a particular decay of a particle called
the long-lived neutral kaon, the $\kl$.  This decay, to a muon and
electron, had been previously credibly ruled out for a branching
fraction (probability per kaon decay) of $10^{-8}$ or higher.  With
newer technology and better beams, the proposal was to search down to
a level of $10^{-12}$.  This decay was forbidden at this level in the
SM, but there was a possibility that the decay occurred at the
$10^{-17}$ level \citep{barroso1984} or lower via a process where
neutrinos change type within an expanded version of the SM; since this
latter process was out of reach, it was included in the ``point null''
hypothesis.  This search was therefore a ``fishing expedition'' for
``physics beyond the Standard Model'' (BSM physics), in this case a
new force of nature with $\sigmatot\approx 10^{-12}$ and $\epsilon_0
\approx 10^{-17}$.  Both the scale $\tau$ of prior belief and
$g(\theta)$ would be hard to define, as the motivation for performing
the experiment was the capability to explore the unknown with the
potential for a major discovery of a new force.  For me personally,
$\pi_1$ was small (say 1\%), and the scale $\tau$ was probably close
to that of the range being explored, $10^{-8}$.  (The first
incarnation of the experiment reached $\sigmatot\approx10^{-11}$,
without evidence for a new force \citep{e791-1993}).  As discussed in
Section~\ref{smalleffect}, searches for such rare decays are typically
interpreted in terms of the influence of possible new particles with
very high masses, higher than can be directly produced.

As another example, perhaps the most extreme, it is of great interest
to determine whether or not protons decay, i.e., whether or not the
decay rate is exactly zero, as so far seems to be the case
experimentally.  Experiments have already probed values of the average
decay rate per proton of 1 decay per $10^{31}$ to $10^{33}$ years.
This is part of the range of values predicted by certain unified field
theories that extend the SM \citep{wilczek2004}.  As the age of the
universe is order $10^{10}$ years, these are indeed very small rates.
Thanks to the exponential nature of such decays in quantum mechanics,
the search for such tiny decay rates is possible by observing nearly
$10^{34}$ protons (many kilotons of water) for several years, rather
than by observing several protons for $10^{34}$ years!  Assigning the
three scales is rather arbitrary, but I would say that
$\sigmatot\approx 10^{-32}$ and $\tau$ initially was perhaps
$10^{-28}$.  Historically the null hypothesis under the SM was
considered to be a point exactly at zero decay rate, until 1976 when
\citet{thooft1976} pointed out an exotic non-perturbative mechanism
for proton decay. But his formula for the SM rate has a factor of
about $\exp(-137\pi) = 10^{-187}$ that makes it negligible even
compared to the BSM rates being explored experimentally. (See
\citet{babu2013} for a recent review.)

Finally, among the multitude of current searches for BSM physics at
the LHC to which Eqn.~\ref{eqn:scales} applies, I mention the example
of the search for production a heavy version of the $\zzero$ boson
(Section~\ref{effectsize}), a so-called $\zp$ (pronounced
``Z-prime'').  The $\zp$ would be the quantum of a new force that
appears generically in many speculative BSM models, but without any
reliable prediction as to whether the mass or production rate is
accessible at the LHC.  For these searches, $\epsilon_0=0$ in the SM;
$\sigmatot$ is determined by the LHC beam energies, intensities, and
the general-purpose detector's measuring capabilities; the scale
$\tau$ is again rather arbitrary (as are $\pi_0$ and $g$), but much
larger than $\sigmatot$.

In all three of these examples, the conditions of
Eqn.~\ref{eqn:scales} are met.  Furthermore, {\em the three scales are
  largely independent}.  There can be a loose connection in that an
experiment may be designed with a particular subjective value of
$\tau$ in mind, which then influences how resources are allocated, if
feasible, to obtain a value of $\sigmatot$ that may settle a
particular scientific issue.  But this kind of connection can be
tenuous in HEP, especially when an existing general-purpose apparatus
such as CMS or ATLAS is applied to a new measurement.  {\em Therefore
  there is no generally applicable rule of thumb relating $\tau$ to
  $\sigmatot$.}

Even if some sense of the scale $\tau$ can be specified, there still
exists the arbitrariness in choosing the form of $g$.  Many
experimenters in HEP think in terms of ``orders of magnitude'', with
an implicit metric that is uniform in the log of the decay rate.  For
example, some might say that ``the experiment is worth doing if it
extends the reach by a factor of 10'', or that ``it is worth taking
data for another year if the number of interactions observed is
doubled''.  But it is not at all clear that such phrasing really
corresponds to a belief that is uniform in the implicit logarithmic
metric.

\subsection{
\texorpdfstring{Test statistics for computing $p$-values in HEP}
                          {Test statistics for computing p-values in HEP}}

\label{heptest} 

There is a long tradition in HEP of using likelihood ratios for both
hypothesis testing and estimation, following established frequentist
theory \citep[Chapter 22]{kendall1999} such as the NP Lemma and
Wilks's Theorem. This is sometimes described in the jargon of HEP
\citep{james1980}, and other times with more extensive sourcing
\citep{eadie1971,bakercousins1984,james2006,ccgv2011}.  When merited,
quite detailed likelihood functions (both binned and unbinned) are
constructed.  In many cases, $\theta$ is a physically non-negative
quantity (such as a mass or a Poisson mean) that vanishes under the
null hypothesis ($\theta_0=0$), and the alternative is $H_1$:
$\theta>0$.  The likelihood-ratio test statistic, denoted by
$\lambda$, and its distribution under the null hypothesis (see below)
are used in a one-tailed test to obtain a $p$-value, which is then
converted to $z$, the equivalent number of standard deviations
($\sigma$) in a one-tailed test of the mean of a normal distribution,
\begin{equation}
\label{eqn:z}
z = \Phi^{-1}(1-p) = \sqrt{2}\, {\rm erf}^{-1}(1-2p).
\end{equation}
For example, $z=3$ corresponds to a $p$-value 
of $1.35 \times 10^{-3}$, and $z=5$ to a $p$-value 
of $2.9 \times  10^{-7}$.
(For upper confidence limits on $\theta$, $p$-values are commonly
modified to mitigate some issues caused by downward fluctuations, 
but this does not affect the procedure for testing $H_0$.)

Nuisance parameters arising from detector calibration, estimates of
background rates, etc., are abundant in these analyses. A large part
of the analysis effort is devoted to understanding and validating the
(often complicated) descriptions of the response of the experimental
apparatus that is included in $\lambda$.  For nuisance parameters, the
uncertainties are typically listed as ``systematic'' in nature, the
name that elementary statistics books use for uncertainties that are
not reduced with more sampling.  Nevertheless, some systematic
uncertainties can be reduced as more data is taken and used in the
subsidiary analyses for calibrations.

A typical example is the calibration of the response of the detector
to a high-energy photon ($\gamma$), crucial for detecting the decay of
the Higgs boson to two photons.  The basic detector response (an
optical flash converted to an analog electrical pulse that is
digitized) must be converted to units of energy.  The resulting energy
``measurement'' suffers from a smearing due to resolution as well as
errors in offset and scale.  Special calibration data and computer
simulations are used to measure both the width and shape of the
smearing function, as well as to determine offsets and scales that
still have residual uncertainty.  In terms of the simple
$N(\theta,\sigmatot^2)$ model discussed throughout this paper, we have
complications: the response function may not be normal but can be
measured; the bias on $\theta$ may not be zero but can be measured;
and $\sigmatot$ is also measured.  All of the calibrations may change
with temperature, position in the detector, radiation damage, etc.
Many resources are put into tracking the time-evolution of calibration
parameters, and therefore minimizing, but of course never eliminating,
the uncertainties.

Such calibration takes place for all the subdetectors used in a HEP
experiment, for all the basic types of detected particles (electrons,
muons, pions, etc.).  Ultimately, with enough data, certain systematic
uncertainties approach constant values that limit the usefulness
adding more data. (Example of limiting systematics would include
finite resolution on the time dependence of detector response; control
of the lasers used for calibration; magnetic field inhomogeneities not
perfectly mapped; imperfect material description in the detector
simulation; and various theoretical uncertainties.)

Once models for the nuisance parameters are selected, various
approaches can be used to ``eliminate'' them from the likelihood ratio
$\lambda$ \citep{cousinsoxford2005}.  ``Profiling'' the nuisances
parameters (i.e., re-optimizing the MLEs of the nuisance parameters
for each trial value of the parameter of interest) has been part of
the basic HEP software tools (though not called profiling) for decades
\citep{james1980}.  The results on the Higgs boson at the LHC have
been based on profiling, partly because asymptotic formulas for
profile likelihoods were generalized \citep{ccgv2011} and found to be
useful.  It is also common to integrate out (marginalize) nuisance
parameters in $\lambda$ in a Bayesian fashion (typically using
evidence-based priors), usually through Monte Carlo integration (while
treating the parameter of interest in a frequentist manner).

In many analyses, the result is fairly robust to the treatment of
nuisance parameters in the definition of $\lambda$. For the separate
step of obtaining the distribution of $\lambda$ under the null
hypothesis, asymptotic theory \citep{ccgv2011} can be applicable, but
when feasible the experimenters also perform Monte Carlo simulations
of pseudo-experiments.  These simulations treat the nuisance
parameters in some frequentist and Bayesian-inspired ways, and are
typically (though not always) rather insensitive to the choice of
method.

To the extent that integrations are performed over the nuisance
parameters, or that profiling yields similar results, the use of
$\lambda$ as a test statistic for a frequentist $p$-value is
reminiscent of Bayesian-frequentist hybrids in the statistics
literature \citep[Section 1]{good1992}, including the prior-predictive
$p$-value of \citet{box1980}.  Within HEP, this mix of paradigms has
been advocated \citep{cousinshighland1992} as a pragmatic approach,
and found in general to yield reasonable results under a variety of
circumstances.

The complexity of such analyses is worth keeping in mind in
Section~\ref{scaletau}, when the concept of the ``unit measurement''
with $\sigma = \sqrt{n}\sigmatot$ is introduced as a basis for some
``objective'' methods of setting the scale $\tau$.  The overall
$\sigmatot$ is a synthesis of many samplings of events of interest as
well as events in the numerous calibration data sets (some disjoint
from the final analysis, some not). {\em It is unclear what could be
  identified as the number of events $n$}, since the analysis does not
fit neatly into the concept of $n$ identical samplings.

\subsection{Are HEP experimenters biased against their null hypotheses?}
Practitioners in disciplines outside of HEP are sometimes accused of
being biased against accepting null hypotheses, to the point that
experiments are set up with the sole purpose of rejecting the null
hypothesis \citep{bayarri1987}.  Strong bias against publishing null
results (i.e., results that do not reject the null hypothesis) has
been described, for example, in psychology \citep{ferguson2012}.
Researchers might feel the need to reject the null hypothesis in order
to publish their results, etc.  It is unclear to what extent these
characterizations might be valid in different fields, but in HEP there
{\em is} often significant prior belief in both the model and the
point null hypothesis (within $\epsilon_0$).  In many searches in HEP,
there is a hope to reject the SM and make a major discovery of BSM
physics in which the SM is nested. But there is nonetheless high (or
certainly non-negligible) prior belief in the null hypothesis.  There
have been hundreds of experimental searches for BSM physics that have
not rejected the SM.

In HEP, it is normal to publish results that advance exploration of
the frontiers even if they do not reject the null hypothesis.  The
literature, including the most prestigious journals, has many papers
beginning with ``Search for\dots'' that report no significant evidence
for the sought-for BSM physics.  Often these publications provide
useful constraints on theoretical speculation, and offer guidance for
future searches.

For physical quantities $\theta$ that cannot have negative values, the
unbiased estimates will be in the unphysical negative region about
half of the time if the true value of $\theta$ is small compared to
$\sigmatot$.  It might appear that the measurement model is wrong if
half the results are unphysical.  But the explanation in retrospect is
that the null hypotheses in HEP have tended to be true, or almost so.
As no BSM physics has been observed thus far at the LHC, the choices
of experiments might be questioned, but they are largely constrained
by resources and by what nature has to offer for discovery.  Huge
detector systems such as CMS and ATLAS are multipurpose experiments
that may not have the desired sensitivity to some specific processes
of interest. Within constraints of available resources and loosely
prioritized as to speculation about where the BSM physics may be
observed, the collaborations try to look wherever there is some
capability for observing new phenomena.

\subsection{Cases of an artificial null that carries little or no belief}

As noted above, the ``core physics models'' used in our searches
typically include the SM as well as larger models in which the SM is
embedded. In a typical search for BSM physics, the SM is the null
hypothesis and carries a non-negligible belief.  However, there does
exist a class of searches for which physicists place {\em little}
prior belief on the null hypothesis, namely when the null hypothesis
is the SM with a missing piece!  This occurs when experimenters are
looking for the ``first observation'' of a phenomenon that {\em is}
predicted by the SM to have non-zero strength $\theta=\theta_1$, but
which is yet to be confirmed in data.  The null hypothesis is then
typically defined to be the simple hypothesis $\theta=\theta_0=0$,
i.e., everything in the SM {\em except} the as-yet-confirmed
phenomenon.  While the alternative hypothesis could be taken to be the
simple hypothesis $\theta=\theta_1$, it is more common to take the
alternative to be $\theta>0$.  Results are then reported in two
pieces: (i) a simple-vs-composite hypothesis test that reports the
$p$-value for the null hypothesis, and (ii) confidence interval(s) for
$\theta$ at one or more confidence level, which can be then compared
to $\theta_1$.  This gives more flexibility in interpretation,
including rejection of $\theta_0=0$, but with a surprising value of
$\hat\theta$ that points to an alternative other than the SM value
$\theta_1$.  Furthermore as in all searches, collaborations typically
present plots showing the distribution of $z$ values obtained from
Monte Carlo simulation of pseudo-experiments under each of the
hypotheses.  From these plots one can read off the ``expected $z$''
(usually defined as median) for each hypothesis, and also get a sense
for how likely is a statistical fluctuation to the observed $z$.

An example from Fermilab is the search for production of single top
quarks via the weak force in proton-antiproton collisions
\citep{d0-singletop,cdf-singletop,PR-singletop}. This search was
performed after the weak force was clearly characterized, and after
top quarks were observed via their production in top-antitop quark
pairs by the strong force.  The search for single top-quark production
was experimentally challenging, and the yields could have differed
from expectations of the \SM due to the possibility of BSM physics.
But there was not much credence in the null hypothesis that production
of single top quarks did not exist at all.  Eventually that null was
rejected at more than 5$\sigma$.  The interest remains on measured
values and particularly confidence intervals for the production rates
(via more than one mechanism), which thus far are consistent with SM
expectations.

Another example is the search for a specific decay mode of the \bs
particle that contains a bottom quark (\quarkb) and anti-strange-quark
(\sbar).  The SM predicts that a few out of $10^9$ \bs decays yield
two muons (heavy versions of electrons) as decay products.  This
measurement has significant potential for discovering BSM physics that
might enhance (or even reduce) the \SM probability for this decay.
The search used the null hypothesis that the \bs decay to two muons
had zero probability, a null that was recently rejected at the
5$\sigma$ level.  As with single top-quark production, the true
physics interest was in the measured confidence interval(s), as there
was negligible prior belief in the artificial null hypothesis of
exactly zero probability for this decay mode. Of course, a
prerequisite for measuring the small decay probability was high
confidence in the presence of this process in the analyzed data. Thus
the clear observation (rejection of the null) at high significance by
each of two experiments was one of the highlights of results from the
LHC in 2013 \citep{cms-bsubs,lhcb-bsubs,PR-bsubs}.

As the Higgs boson is an integral part of the SM (required for the
renormalizability of the SM) , the operational null hypothesis used in
searching for it was similarly taken to be an artificial model that
included all of the SM except the Higgs boson, and which had no BSM
physics to replace the Higgs boson with a ``Higgs-like'' boson.
However, the attitude toward the hypotheses was not as simple as in
the two previous examples.  The null hypothesis of having ``no Higgs
boson'' carried some prior belief, in the sense that it was perhaps
plausible that BSM physics might mean that no SM Higgs boson (or
Higgs-like boson) was observable in the manner in which we were
searching.  Furthermore, the search for the Higgs boson had such a
long history, and had become so well-known in the press, that there
would have been a notable cost to a false discovery claim.  In my
opinion, this was an important part of the justification for the high
threshold that the experimenters used for declaring an observation.
(Section~\ref{npalpha} discusses factors affecting the threshold.)

Analogous to the two previous examples, the implementation of the
alternative hypothesis was as the complete SM with a composite
$\theta$ for the strength of the Higgs boson signal.  (This
generalized alternative allowed for a ``Higgs-like'' boson that
perhaps could not be easily distinguished with data in hand.)
However, the mass of the Higgs boson is a free parameter in the SM,
and had been only partially constrained by previous measurements and
theoretical arguments.  Compared to the two previous examples, this
complicated the search significantly, as the probabilities of
different decay modes of the Higgs boson change dramatically as a
function of its mass.

This null hypothesis of no Higgs (or Higgs-like) boson was
definitively rejected upon the announcement of the observation of a
new boson by both ATLAS and CMS on July 4, 2012. The confidence
intervals for signal strength $\theta$ in various decay sub-classes,
though not yet precise, were in reasonable agreement with the
predictions for the SM Higgs boson. Subsequently, much of the focus
shifted to measurements of describing different production and decay
mechanisms.  For measurements of continuous parameters, the null
hypothesis has reverted to the complete SM with its Higgs boson, and
the tests (e.g., \citet[Figure 22]{cmshiggsprop4lep2013} and
\citet[Figures 10-13]{atlashiggsprop2013}) use the frequentist duality
(Section~\ref{npalpha} below) between interval estimation and
hypothesis testing.  One constructs (approximate) confidence intervals
and regions for parameters controlling various distributions, and
checks whether the predicted values for the SM Higgs boson are within
the confidence regions. For an important simple-vs-simple hypothesis
test of the quantum mechanical property called parity, $p$-values for
both hypotheses were reported \citep{cmshiggsprop2012}, as described
in Section~\ref{simple}.

\section{\texorpdfstring{\boldmath What sets the scale $\tau$?}
              {What sets the scale tau?}}
\label{scaletau}

As discussed by \citet[p.~251]{jeffreys1961} and re-emphasized by
\citet{bartlett1957}, defining the scale $\tau$ (the range of values
of $\theta$ over which the prior $g(\theta)$ is relatively large) is a
significant issue. Fundamentally, the scale appears to be personal and
subjective, as is the more detailed specification of $g(\theta)$.
\citet{bergerdelampady1987,{bergerdelampady1987r}} state that ``the
precise null testing situation is a prime example in which objective
procedures do not exist,'' and ``Testing a precise hypothesis is a
situation in which there is {\em clearly} no objective Bayesian
analysis and, by implication, no sensible objective analysis
whatsoever.''  Nonetheless, as discussed in this section, Berger and
others have attempted to formulate principles for specifying default
values of $\tau$ for communicating scientific results.

\citet{bartlett1957} suggests that $\tau$ might scale as $1/\sqrt{n}$,
canceling the sample-size scaling in $\sigmatot$ and making the Bayes
factor independent of $n$. \citet[p.~106]{cox2006} suggests this as
well, on the grounds that ``\dots{}in most, if not all, specific
applications in which a test of such a hypothesis [$\theta=\theta_0$]
is thought worth doing, the only serious possibilities needing
consideration are that either the null hypothesis is (very nearly)
true or that some alternative within a range fairly close to
$\theta_0$ is true.''  This avoids the situation that he finds
unrealistic, in which ``the corresponding answer depends explicitly on
$n$ because, typically unrealistically, large portions of prior
probability are in regions remote from the null hypothesis relative to
the information in the data.''  Part of Cox's argument was already
given by \citet[p.~251]{jeffreys1961}, ``\dots{}the mere fact that it
has been suggested that [$\theta$] is zero corresponds to some
presumption that [$\theta$] is small.''  \citet[p.~114]{leamer1978}
makes a similar point, ``\dots{}a prior that allocates positive
probability to subspaces of the parameter space but is otherwise
diffuse represents a peculiar and unlikely blend of knowledge and
ignorance''.  (As Section~\ref{hepex} discusses, this ``peculiar and
unlikely blend'' is common in HEP.) \citet{andrews1994} also explores
the consequences of $\tau$ shrinking with sample size, but these ideas
seem not to have led to a standard.  As another possible
reconciliation, \citet{robert1993} considers $\pi_1$ that increases
with $\tau$, but this seems not to have been pursued further.

Many attempts in the Bayesian literature to specify a default $\tau$
arrive at a suggestion that does {\em not} depend on $n$, and hence
does not remove the dependence of the Ockham factor on $n$.  In the
search for any non-subjective $n$-independent scale, the only option
seemingly at hand is $\sigmatot$ when $n=1$, i.e., the original
$\sigma$ (Eqn.~\ref{eqn:sigmatot}) that expresses the uncertainty of a
single measurement.  This was in fact suggested by
\citet[p.~268]{jeffreys1961}, on the grounds that there is nothing
else in the problem that can set the scale, and was followed, for
example, in generalizations by \citet*{zellnersiow1980}.

\citet*{kasswasserman1995} do the same, which ``has the interpretation
of `the amount of information in the prior on [$\theta$] is equal to
the amount of information about [$\theta$] contained in one
observation' ''.  They refer to this as a ``unit information prior'',
citing \citet{smithspiegelhalter1980} as also using this ``appealing
interpretation of the prior.''  It is not clear to me why this ``unit
information'' approach is ``appealing'', or how it could lead to
useful, universally cross-calibrated Bayes factors in HEP.  As
discussed in Section~\ref{heptest} the detector may also have some
intrinsic $\sigmatot$ for which no preferred $n$ is evident.
\citet[pp.~132, 135]{raftery1995} points out the same problem.  After
defining a prior for which, ``roughly speaking, the prior distribution
contains the same amount of information as would, on average, one
observation'', he notes the obvious problem in practice: the
``important ambiguity\dots{}the definition of [$n$], the sample
size.'' He gives several examples for which he has a recommendation.

\citet[with commentary]{bergerpericchi2001} review more general
possibilities based on use of the information in a small subset of the
data, and for one method claim that ``this is the first general
approach to the construction of conventional priors in nested
models.''  \citet{berger2007phystat,berger2011phystat} applied one of
these so-called ``intrinsic priors'' to a pedagogical example and its
generalization from HEP.  Unfortunately, I am not
aware of anyone in HEP who has pursued these suggestions.  Meanwhile,
recently \citet*{bayarri2012} have reconsidered the issue and
formulated principles resulting ``\dots{}in a new model selection
objective prior with a number of compelling properties.''  I think
that it is fair to conclude that this is still an active area of
research.

\subsection{Comments on non-subjective priors for estimation 
and model selection}

For {\em point and interval estimation}, \citet{jeffreys1961} suggests
two approaches for obtaining a prior for a physically non-negative
quantity such as the magnitude of the charge $q$ of the electron.
Both involve invariance concepts.  The {\em first} approach
(pp.~120-123) considers only the parameter being measured.  In his
example, one person might consider the charge $q$ to be the
fundamental quantity, while another might consider $q^2$ (or some
other power $q^m$ ) to be the fundamental quantity.  In spite of this
arbitrariness of the power $m$, everyone will arrive at consistent
posterior densities if they each take the prior for $q^m$ to be
$1/q^m$, since all expressions $d(q^m)/q^m)$ differ only by a
proportionality constant.  (Equivalently, they can all take the prior
as uniform in $\ln q^m$, i.e., in $\ln q$.)

Jeffreys's more famous {\em second} approach, leading to his eponymous
rule and priors, is based on the likelihood function and some averages
over the sample space (i.e., over possible observations).  The
likelihood function is based on what statisticians call the
measurement ``model''.  This means that ``Jeffreys's prior'' is
derived {\em not} by considering only the parameter being measured,
but rather by examining the {\em measuring apparatus}.  For example,
Jeffreys's prior for a Gaussian (normal) measurement apparatus is
uniform in the measured value.  If the measuring apparatus has
Gaussian response in $q$, the prior is uniform in $q$.  If the
measuring apparatus has Gaussian response in $q^2$, then the prior is
uniform in $q^2$.  If the physical parameter is measured with Gaussian
resolution and is physically non-negative, as for the charge magnitude
$q$, then the functional form of the prior remains the same (uniform)
and is set to zero in the unphysical region
\citep[p.~89]{bergerdecision1985}.

Berger and Bernardo refer to ``non-subjective'' priors such as
Jeffreys's prior as ``objective'' priors.  This strikes me as rather
like referring to ``non-cubical'' volumes as ``spherical'' volumes;
one is changing the usual meaning to the word.
\citet{bernardo2011bayes} defends the use of ``objective'' as follows.
``No statistical analysis is really objective, since both the
experimental design and the model assumed have very strong subjective
inputs.  However, frequentist procedures are often branded as
`objective' just because their conclusions are only conditional on the
model assumed and the data obtained.  Bayesian methods where the prior
function is directly derived from the assumed model are objective in
this limited, but precise sense.''

Whether or not this defense is accepted, so-called ``objective''
priors can be deemed useful for point and interval {\em estimation},
even to frequentists, as there is a deep (frequentist) reason for
their potential appeal.  Because the priors are derived by using
knowledge of the properties of the {\em measuring apparatus}, it is at
least conceivable that Bayesian credible intervals based on them might
have better-than-typical frequentist coverage properties when
interpreted as approximate frequentist confidence intervals.  As
\citet{welchpeers1963} showed, for Jeffreys's priors this is indeed
the case for one-parameter problems. Under suitable regularity
conditions, the approximate coverage of the resulting Bayesian
credible intervals is uniquely good to order $1/n$, compared to the
slower convergence for other priors, which is good to order
$1/\sqrt{n}$.  Hence, except at very small $n$, by using ``objective''
priors, one can (at least approximately) obey the Likelihood Principle
and obtain decent frequentist coverage, which for some is a preferred
``compromise''.  Reasonable coverage can also be the experience for
Reference Priors with more than one parameter \citep[and references
therein]{philippe1998}.  This can happen even though objective priors
are improper (i.e., not normalizable) for many prototype problems; the
ill-defined normalization constant cancels out in the calculation of
the posterior. (Equivalently, if a cutoff parameter is introduced to
make the prior proper, the dependence on the cutoff vanishes as it
increases without bound.)

For model selection, Jeffreys proposed a {\em third} approach to
priors.  As discussed in Sections~\ref{original} and \ref{pointnull},
from the point of view of the larger model, the prior is irregular, as
it is described by a probability mass (a Dirac $\delta$-function) on
the null value $\theta_0$ that has measure zero.  The prior
$g(\theta)$ on the rest of $\Theta$ must be normalizable (eliminating
improper priors used for estimation) in order for the posterior
probability to be well-defined.  For Gaussian measurements, Jeffreys
argued that $g$ should be a Cauchy density (``Breit-Wigner'' in HEP).

Apart from the subtleties that led Jeffreys to choose the Cauchy form
for $g$, there is the major issue of the scale $\tau$ of $g$, as
discussed in Section~\ref{scaletau}.  The typical assumption of
``objective Bayesians'' is that, basically by definition, an objective
$\tau$ is one that is derived from the measuring apparatus.  And then,
{\em assuming that $\sigmatot^2$ reflects $n$ measurements using an
  apparatus that provides a variance for each of $\sigma^2$, as in
  Eqn.~\ref{eqn:sigmatot}}, they invoke $\sigma$ as the scale of the
prior $g$.

Lindley (e.g., in commenting on \citet{bernardo2011bayes}) argues in
cases like this that objective Bayesians can get lost in the Greek
letters and lose contact with the actual context.  I too find it
puzzling that one can first argue that the Ockham's factor is a
crucial feature of Bayesian logic that is absent from frequentist
reasoning, and then resort to choosing this factor based on the
measurement apparatus, and on a concept of sample size $n$ that can be
difficult to identify.  The textbook by \citet[p.~130]{lee2004}
appears to agree that this is without compelling foundation:
``Although it seems reasonable that [$\tau$] should be chosen
proportional to [$\sigma$], there does not seem to be any convincing
argument for choosing this to have any particular value\dots.''

It seems that, in order to be useful, any ``objective'' choice of
$\tau$ must provide demonstrable cross-calibration of experiments with
different $\sigmatot$ when $n$ is not well-defined.  Another voice
emphasizing the practical nature of the problem is that of
\citet{kass2009}, saying that Bayes factors for hypothesis testing
``remain sensitive---to first order---to the choice of the prior on
the parameter being tested.''  The results are ``contaminated by a
constant that does not go away asymptotically.''  He says that this
approach is ``essentially nonexistent'' in neuroscience.

\section{The reference analysis approach of Bernardo}
\label{bernardo}

\citet{bernardo1999bayes} (with critical discussion) defines Bayesian
hypothesis testing in terms very different from calculating the
posterior probability of $H_0$: $\theta=\theta_0$. He proposes to
judge whether $H_0$ is {\em compatible} (his italics) with the data:

``Any Bayesian solution to the problem posed will obviously require a
prior distribution $p(\theta)$ over $\Theta$, and the result may well
be very sensitive to the particular choice of such prior; note that,
in principle, there is no reason to assume that the prior should
necessarily be concentrated around a particular $\theta_0$; indeed,
for a judgement on the compatibility of a particular parameter value
with the observed data to be useful for scientific communication, this
should only depend on the assumed model and the observed data, and
this requires some form of non-subjective prior specification for
$\theta$ which could be argued to be `neutral'; a sharply concentrated
prior around a particular $\theta_0$ would hardly qualify.''  He later
continues, ``\dots{}nested hypothesis testing problems are better
described as specific decision problems about the choice of a useful
model and that, when formulated within the framework of decision
theory, they do have a natural, fully Bayesian, coherent solution.''

Unlike Jeffreys, Bernardo advocates using the {\em same}
non-subjective priors (even when improper) for hypothesis testing as
for point and interval estimation.  He defines a discrepancy measure
$d$ whose scaling properties can be complicated for small $n$, but
which asymptotically can be much more akin to those of $p$-values than
to those of Bayes factors.  In fact, if the posterior becomes
asymptotically normal, then $d$ approaches $(1+z^2)/2$
\citep{bernardo2011phystat,bernardo2011bayes}.  A fixed cutoff for his
$d$ (which he regards as the objective approach), just as a fixed
cutoff for $z$, is {\em inconsistent} in the statistical sense, namely
it does not accept $H_0$ with probability 1 when $H_0$ is true and the
sample size increases without bound.

\citet{bernardorueda2002} elaborate this approach further, emphasizing
that the Bayes factor approach, when viewed from the framework of
Bernardo's formulation in terms of decision theory, corresponds to a
``zero-one'' loss-difference function, which they refer to as
``simplistic''.  (Loss functions are discussed by \citet[Section
2.4]{bergerdecision1985}.)  The zero-one loss is so-named because the
loss is zero if a correct decision is made, and 1 if an incorrect
decision is made. Berger states that, in practice, this loss will
rarely be a good approximation to the true loss.)  Bernardo and Rueda
prefer continuous loss functions (such as quadratic loss) that do not
require the use of non-regular priors.  A prior sharply spiked at
$\theta_0$ ``{\em assumes} important prior knowledge \dots{\em very
  strong} prior beliefs,'' and hence ``Bayes factors should {\em not}
be used to test the {\em compatibility} of the data with $H_0$, for
they inextricably combine what the data have to say with (typically
subjective) {\em strong} beliefs about the value of $\theta$.'' This
contrasts with the commonly followed statement of
\citet[p.~246]{jeffreys1961} that (in present notation), ``To say that
we have no information initially as to whether the new parameter is
needed or not we must take $\pi_0 = \pi_1 = 1/2$''. Bernardo and Rueda
reiterate Bernardo's above-mentioned recommendation of applying the
discrepancy measure (expressed in ``natural'' units of information)
according an {\em absolute} scale that is independent of the specific
problem.

\citet{bernardo2011bayes} provides a major review (with extensive
commentary), referring unapprovingly to point null hypotheses in an
``objective'' framework, and to the use begun by Jeffreys of two
``{\em radically different}{}'' types of priors for estimation and for
hypothesis testing.  He clarifies his view of hypothesis testing, that
it is a decision whether ``to act {\em as if} $H_0$ were true'', based
on the expected posterior loss from using the simpler model rather
than the alternative (full model) in which it is nested.

In his rejoinder, Bernardo states that the \JL paradox ``clearly poses
a very serious problem to Bayes factors, in that, under certain
conditions, they may lead to misleading answers. Whether you call this
a paradox or a disagreement, the fact that the Bayes factor for the
null may be arbitrarily large for sufficiently large $n$, {\em however
  relatively unlikely the data may be under} $H_0$ is, to say the
least, deeply disturbing\dots the Bayes factor analysis may be
completely misleading, in that it would suggest {\em accepting} the
null, even if the likelihood ratio for the MLE {\em against} the null
is very large.''

At a recent PhyStat workshop where \citet{bernardo2011phystat}
summarized this approach, physicist \citet{demortier2011phystat}
considered it appropriate when the point null hypothesis is a {\em
  useful} simplification (in the sense of definitions in decision
theory) rather a point having significant prior probability. He noted
(as did Bernardo) that the formalism can account for point nulls if
this is desired.

\section{Effect size in HEP}
\label{effectsize}
As noted in the introduction, in this paper ``effect size'' refers to
the point and interval estimates (measured values and uncertainties)
of a parameter or physical quantity, typically expressed in the
original units.  Apparently, reporting of effect sizes is not always
automatic in some disciplines, leading to repeated reminders to report
them \citep{kirk1996,wilkinson1999,nakagawa2007,apa2010}.  In HEP,
however, point estimates and confidence intervals for model parameters
are used to summarize the results of nearly all experiments, and to
compare to the predictions of theory (which often have uncertainties
as well).

For experiments in which one particle interacts with another, the
meeting point for comparison of theory and experiment is frequently an
interaction probability referred to as a ``cross section''.
%
%
For particles produced in interactions and that subsequently decay
(into other particles), the comparison of theory and experiment
typically involves the decay rate (probability of decay per second)
or its inverse, the mean lifetime.
Measurements of cross sections and decay rates can be subdivided into
distinguishable subprocesses, as functions of both continuous
parameters (such as production angles) and discrete parameters (such
as the probabilities known as ``branching fractions'' for decay into
different sets of decay products).

In the example of the Higgs boson discovery, the effect size was
quantified through confidence intervals on the product of cross
sections and the branching fractions for different sets of decay
products.  These confidence intervals provided exciting indications
that the new boson was indeed ``Higgs-like'', as described by
Incandela and Gianotti and the subsequent ATLAS and CMS publications
\citep{atlashiggs,cmshiggs}.  By spring 2013, more data had been
analyzed and it seemed clear to both collaborations that the boson was
``a'' Higgs boson (leaving open the possibility that there might be
more than one).  Some of the key figures are described in the
information accompanying the announcement of the 2013 Nobel Prize in
Physics \citep[Figures 6 and 7]{nobel2013}.

\subsection{No effect size is too small in core physics models}

If one takes the point of view that ``all models are wrong''
\citep{box1976}, then a tiny departure from the null hypothesis for a
parameter in a normal model, which is conditional on the model being
true, might be properly disregarded as uninteresting.  Even if the
model is true, a small $p$-value might be associated with a departure
from the null hypothesis (effect size) that is too small to have
practical significance in formulating public policy or
decision-making.  In contrast, core physics models reflect presumed
``laws of nature'', and it is always of major interest if departures
with any effect size can be established with high confidence.

In HEP, tests of core physics models also benefit from what we believe
to be the world's most perfect random-sampling mechanism, namely
quantum mechanics.  In each of many repetitions of a given initial
state, nature randomly picks out a final state according to the
weights given by the (true, but not completely known) laws of physics
and quantum mechanics.  Furthermore, the most perfect incarnation of
``identical'' is achieved through the fundamental quantum-mechanical
property that elementary particles of the same type are {\em
  indistinguishable}.  The underlying statistical model is typically
binomial or its generalizations and approximations, especially the
Poisson distribution.

\subsection{Small effect size can indicate new phenomena at higher energy}
\label{smalleffect}
For every force there is a quantum field that permeates all space.  As
suggested in 1905 by Einstein for the electromagnetic (EM) field,
associated with every quantum field is an ``energy quantum'' (called
the photon for the EM field) that is absorbed or emitted
(``exchanged'') by other particles interacting via that field.  While
the mass of the photon is presumed to be exactly zero, the masses of
quanta of some other fields are non-zero.  The nominal mass $m$,
energy $E$, and momentum $p$ of such energy quanta are related through
Einstein's equation, $m^2 = E^2 - p^2$.  (For unstable particles, the
definition of the nominal mass is somewhat technical, but there are
agreed-on conventions.)

Interactions in modern physics are possible because energy quanta can
be exchanged even when the energy $\Delta E$ and momentum $\Delta p$
being transferred in the interaction do not correspond to the nominal
mass of the exchanged quantum.  With a quantity $q^2$ (unrelated to
symbol for the charge $q$ of the electron) defined by $q^2 = (\Delta
E)^2 - (\Delta p)^2$, quantum mechanics reduces the probability of the
reaction as $q^2$ departs from the true $m^2$ of the exchanged
particle.  In many processes, the reduction factor is at leading order
proportional to
\begin{equation}
\label{virtual}
\frac{1}{(m^2 - q^2)^2}.
\end{equation}
(As $q^2$ can be positive or negative, the relative sign of $q^2$ and
$m^2$ depends on details of the process. For positive $q^2$, the
singularity of $m^2=q^2$ is made finite by another term that can be
otherwise neglected in the present discussion.)  What $q^2$ is
accessible depends on the available technology; in general, larger
$q^2$ requires higher-energy particle beams and therefore more costly
accelerators.

For the photon, $m=0$, and the interaction probability goes as
$1/q^4$.  On the other hand, if the mass $m$ of the quantum of a force
is so large that $m^2\gg|q^2|$, then the probability for an
interaction to occur due to the exchange of such a quantum is
proportional to $1/m^4$.  {\em By looking for interactions or decays
  having {\em very} low probability, it is possible to probe the
  existence of massive quanta with $m^2$ well beyond those that can be
  created with concurrent technology.}

An illustrative example, studied by \citet{galison1983}, is the
accumulation of evidence for the $\zzero$ boson (with mass $\mz$), an
electrically neutral quantum of the weak force hypothesized in the
1960s.  Experiments were performed in the late 1960s and early 1970s
using intense beams of neutrinos scattering off targets of ordinary
matter.  The available $|q^2|$ was much smaller than $\mz^2$,
resulting in a small reaction probability in the presence of other
processes that obscured the signal.  CERN staked the initial claim for
observation \citep{hasert1973}. After a period of confusion, both CERN
and Fermilab experimental teams agreed that they had observed
interactions mediated by $\zzero$ bosons, even though no $\zzero$
bosons were detected directly, as the energies involved (and hence
$\sqrt{|q^2|}$) were well below $\mz$.

In another type of experiment probing the $\zzero$ boson, conducted at
SLAC in the late 1970s \citep{prescott1978}, specially prepared
electrons (``spin polarized electrons'' in physics jargon) were
scattered off nuclei to seek a very subtle left-right asymmetry in the
scattered electrons arising from the combined action of
electromagnetic and weak forces.  In an exquisite experiment, an
asymmetry of about 1 part in $10^4$ was measured to about 10\%
statistical precision with an estimated systematic uncertainty also
about 10\%.  The statistical model was binomial, and the experiment
had the ability to measure departures from unity of twice the binomial
parameter with an uncertainty of about $10^{-5}$.  I.e., the sample
size of scattered electrons was of order $10^{10}$.  This precision in
a binomial parameter is finer than that in an ESP example that has
generated lively discussion in the statistics literature on the \JL
paradox \citep[pp.\ 19, 26, and cited references, and comments and
rejoinder]{bernardo2011bayes}.  More recent experiments measure this
scattering asymmetry even more precisely. The results of Prescott et
al.\ confirmed predictions of the model of electroweak interactions
put forward by Glashow, Weinberg, and Salam, clearing the way for
their Nobel Prize in 1979.

Finally, in 1982, the technology for creating interactions with
$q^2=\mz^2$ was realized at CERN through collisions of high energy
protons and antiprotons (and subsequently at Fermilab). And in 1989,
``\zzero\ factories'' turned on at SLAC and CERN, colliding electrons
and positrons at beam energies tuned to $q^2=\mz^2$. At this $q^2$,
the small denominator in Eqn.~\ref{virtual} causes the {\em tiny}
deviation in the previous experiments to become a {\em huge} increase
in the interaction probability, a factor of 1000 increase compared to
the null hypothesis of ``no $\zzero$ boson''.  (There is an additional
term in the denominator of Eqn.~\ref{virtual} that reflects the
instability of the $\zzero$ boson to decay and that I have neglected
thus far; at $q^2=\mz^2$, it keeps the expression finite.)

This sequence of events in the experimental pursuit of the $\zzero$
boson is somewhat of a prototype for what many in HEP hope will happen
again.  A given process (scattering or decay) has rate zero (or
immeasurably small $\epsilon_0$) according to the SM.  If, however,
there is a new boson $X$ with mass $m_X$ much higher than accessible
with current technology, then the boson may give a non-zero rate,
proportional to $1/m_X^4$, for the given process.  The null hypothesis
is that $X$ does not exist and the rate for the process is
immeasurably small.  As $m_X$ is not known, the possible rates for the
process if $X$ {\em does} exist comprise a continuum, including rates
arbitrarily close to zero.  But these tiny numbers in the continuum
map onto possibilities for major, discrete, modifications to the laws
of nature---new forces!

The searches for rare decays described in Section~\ref{hepex} are
examples of this approach.  For rare decays of $\kl$ particles, an
observation of a branching fraction at the $10^{-11}$ level would have
indicated the presence of a new mass scale some 1000 times greater
than the mass of the $\zzero$ boson, which is more than a factor of 10
above currently accessible $q^2$ values at LHC.  Such mass scales are
also probed by measuring the difference between the mass of the $\kl$
and that of closely related particle, the short-lived neutral kaon
($\ks$).  The mass of the $\kl$ is about half the mass of the proton,
and has been measured to a part in $10^4$.  The $\kl-\ks$ mass
difference has been measured to a part in $10^{14}$, far more
precisely than the mass itself.  The difference arises from
higher-order terms in the weak interaction, and is extremely sensitive
to certain classes of speculative BSM physics.  Even more
impressively, the observation of proton decay with a decay rate at the
level probed by current experiments would spectacularly indicate a new
mass scale a factor of $10^{13}$ greater than that of the mass of the
$\zzero$ boson.

Alas, none of these experiments has observed processes that would
indicate BSM physics.  In the intervening years, there have however
been major discoveries in neutrino physics that have redefined and
extended the SM.  These discoveries established that the mass of the
neutrino, while tiny, is not zero.  In some physics models called
``seesaw models'', the neutrino mass is inversely proportional to a
mass scale of BSM physics; thus one interpretation is that the tiny
neutrino masses indicate a new very large mass scale, perhaps
approaching the scale probed by proton decay \citep{sciam2013}.

\section{\texorpdfstring
{\boldmath Neyman-Pearson testing and the choice of \\ Type I error probability $\alpha$}
{Neyman-Pearson testing and the choice of Type I error probability alpha}}
\label{npalpha}

In Neyman-Pearson (NP) hypothesis testing, the Type I error $\alpha$
is the probability of rejecting $H_0$ when it is true.  For testing a
point null vs a composite alternative, there is a duality between NP
hypothesis testing and frequentist interval estimation via confidence
intervals.  The hypothesis test for $H_0$: $\theta=\theta_0$ vs $H_1$:
$\theta\ne\theta_0$, at significance level (``size'') $\alpha$, is
entirely equivalent to whether $\theta_0$ is contained in a confidence
interval for $\theta$ with confidence level (CL) of $1-\alpha$.  As
emphasized by \citet*[p.~175]{kendall1999}, ``Thus there is no need to
derive optimal properties separately for tests and intervals: there is
a one-to-one correspondence between the problems\dots.''

\citet{mayospanos2006} argue that confidence intervals have
shortcomings that are avoided by using Mayo's concept of ``severe
testing''.  \citet{spanos2013} argues this specifically in the context
of the \JL paradox.  I am not aware of widespread application of the
severe testing approach, and do not yet understand it well enough to
see how it would improve scientific communication in HEP if adopted.
Hence the present paper focuses on the traditional frequentist
methods.

As mentioned in Section~\ref{heptest}, in HEP the workhorse test
statistic for testing and estimation is often a likelihood-ratio
$\lambda$.  In practice, sometimes one first performs the hypothesis
test and uses the duality to ``invert the test'' to obtain confidence
intervals, and sometimes one first finds intervals.  Performing the
test and inverting it in a rigorous manner is equivalent to the
original ``Neyman construction'' of confidence intervals
\citep{neyman1937}.  Such a construction using the likelihood-ratio
test statistic has been advocated by \citet{feldman1998}, particularly
in irregular problems such as when the null hypothesis is on the
boundary.  In more routine applications, approximate confidence
intervals or regions can be obtained by finding maximum-likelihood
estimates of unknown parameters and forming regions bounded by
contours of differences in $\ln\lambda$ as in Wilks's Theorem
\citep{james1980,james2006}.

Confidence intervals in HEP are typically presented for conventional
confidence levels (68\%, 90\%, 95\%, etc.).  Alternatively, when
experimenters report a $p$-value with respect to some null value,
anyone can invoke the \NP accept/reject paradigm by comparing the
reported $p$-value to one's own (previously chosen) value of $\alpha$.
From a mathematical point of view, one can {\em define} the post-data
$p$-value as the smallest significance level $\alpha$ at which the
null hypothesis would be rejected, had that $\alpha$ been specified in
advance \citep[p.~335]{rice2007}.  This may offend some who point out
that Fisher did not define the $p$-value this way when he introduced
the term, but these protests do not negate the numerical identity with
Fisher's $p$-value, even when the different interpretations are kept
distinct.

Regardless of the steps through which one learns whether the test
statistic $\lambda$ is in the rejection region of a particular value
of $\theta$, one must choose the size $\alpha$, the Type I error
probability of rejecting $H_0$ when it is true.  Neyman and Pearson
introduced the alternative hypothesis $H_1$ and the Type II error
$\beta$ for the probability under $H_1$ that $H_0$ is not rejected
when it is false.  They remarked, \citep[p.~296]{neymanpearson1933a}
``These two sources of error can rarely be eliminated completely; in
some cases it will be more important to avoid the first, in others the
second. \dots{}The use of these statistical tools in any given case,
in determining just how the balance should be struck, must be left to
the investigator.''

\citet*[p.~57, and earlier editions by Lehmann]{lehmann2005} echo this
point in terms of the {\em power} of the test, defined as $1-\beta$:
``The choice of a level of significance $\alpha$ is usually somewhat
arbitrary\dots{}the choice should also take in consideration the power
that the test will achieve against the alternatives of
interest\dots.''

For simple-vs-simple hypothesis tests discussed in
Section~\ref{simple}, the power 1$-${}$\beta$ is well-defined, and, in
fact, \citet[p.~497]{neymanpearson1933b} discuss how to balance the
two types of error, for example by considering their sum.  It is
well-known today that such an approach, including minimizing a
weighted sum, can remove some of the unpleasant aspects of testing
with a fixed $\alpha$, such as inconsistency in the statistical sense
(as mentioned in Section~\ref{bernardo}, not accepting $H_0$ with
probability 1 when $H_0$ is true and the sample size increases without
bound).

But this optimization of the tradeoff between $\alpha$ and $\beta$
becomes ill-defined for a test of simple vs composite hypotheses when
the composite hypothesis has values of $\theta$ arbitrarily close to
$\theta_0$, since the limiting value of $\beta$ is 0.5, independent of
$\alpha$ \citep[p.~496]{neymanpearson1933b}. \citet{robert2013} echoes
this concern that in NP testing, ``there is a fundamental difficulty
in finding a proper balance (or imbalance) between Type I and Type II
errors, since such balance is not provided by the theory, which
settles for the sub-optimal selection of a {\em fixed} Type I error.
In addition, the whole notion of {\em power}, while central to this
referential, has arguable foundations in that this is a {\em function}
that inevitably depends on the unknown parameter $\theta$.  In
particular, the power decreases to the Type I error at the boundary
between the null and the alternative hypotheses in the parameter
set.''

Unless a value of $\theta$ in the composite hypothesis is of
sufficiently special interest to justify its use for considering
power, there is no clear procedure.  A Bayesian-inspired approach
would allow optimization by weighting the values of $\theta$ under
$H_1$ by a prior $g(\theta)$.  As \citet[p.~142]{raftery1995} notes,
``Bayes factors can be viewed as a precise way of implementing the
advice of [\citet{neymanpearson1933a}] that power and significance be
balanced when setting the significance level\dots{}there is a conflict
between Bayes factors and significance testing at predetermined levels
such as .05 or .01.''  In fact, \citet[p.~502]{neymanpearson1933b}
suggest this possibility if multiple $\theta_i$ under the alternative
hypothesis are genuinely sampled from known probabilities $\Phi_i$:
``\dots{}if the $\Phi_i$'s were known, a test of greater resultant
power could almost certainly be found.''

Kendall and Stuart and successors \citep*[Section 20.29]{kendall1999}
view the choice of $\alpha$ in terms of costs: ``\dots{}unless we have
supplemental information in the form of the {\em costs} (in money or
other common terms) of the two types of error, and costs of
observations, we cannot obtain an optimal combination of $\alpha$,
$\beta$, and $n$ for any given problem.''  But prior belief should
also play a role, as remarked by \citet*[p.~58]{lehmann2005} (and
earlier editions by Lehmann): ``Another consideration that may enter
into the specification of a significance level is the attitude toward
the hypothesis before the experiment is performed.  If one firmly
believes the hypothesis to be true, extremely convincing evidence will
be required before one is willing to give up this belief, and the
significance level will accordingly be set very low.''

Of course, these vague statements about choosing $\alpha$ do not come
close to a formal decision theory (which is however not visibly
practiced in HEP).  For the case of simple vs composite hypotheses
relevant to the \JL paradox, HEP physicists informally take into
account prior belief, the measured values of $\theta$ and its
confidence interval, as well as relative costs of errors, contrary to
myths about automatic use of a ``5$\sigma$'' criterion discussed in
the next section.

\subsection{\texorpdfstring{The mythology of 5$\sigma$}
{The mythology of five sigma}}
\label{fivesigma} 
Nowadays it is commonly written that 5$\sigma$ is the criterion for a
discovery in HEP.  Such a fixed one-size-fits-all
level of significance ignores the consideration noted above by
Lehmann, and violates one of the most commonly stated tenets of
science---that the more extraordinary the claim, the more
extraordinary must be the evidence.  I do not believe that experienced
physicists have such an automatic response to a $p$-value, but it may
be that some people in the field may take the fixed threshold more
seriously than is warranted.

The (quite sensible) historical roots of the 5$\sigma$ criterion were
in a specific context, namely searches performed in the 1960s for new
``elementary particles'', now known to be composite particles with
different configurations of quarks in their substructure.  A plethora
of histograms were made, and presumed new particles, known as
``resonances'' showed up as localized excesses (``bumps'') spanning
several histogram bins.  Upon finding an excess and defining those
bins as the ``signal region'', the ``local $p$-value'' could be
calculated as follows.  First the nearby bins in the histogram
(``sidebands'') were used to formulate the null hypothesis
corresponding to the expected number of events in the signal region in
the absence of a new particle.  Then the (Poisson) probability under
the null hypothesis of observing a bump as large as or larger than
that seen was calculated, and expressed in terms of standard
deviations ``$\sigma$'' by analogy to a one-sided test of a normal
distribution.

The problem was that the location of a new resonance was typically not
known in advance, and the local $p$-value did not include the fact
that ``pure chance'' had lots of opportunities (lots of histograms and
many bins) to provide an unlikely occurrence.  Over time many of the
alleged new resonances were not confirmed in other independent
experiments.  In the group led by Alvarez at Berkeley, histograms with
putative new resonances were compared to simulations drawn from smooth
distributions \citep{alvarez1968}.  \citet[p.~465]{rosenfeld1968}
describes such simulations and rough hand calculations of the number
of trials, and concludes, ``To the theorist or phenomenologist the
moral is simple: wait for nearly 5$\sigma$ effects. For the
experimental group who have spent a year of their time and perhaps a
million dollars, the problem is harder\dots{}go ahead and
publish\dots{}but they should realize that any bump less than about
5$\sigma$ calls only for a repeat of the experiment.''

The original concept of ``5$\sigma$'' in HEP was therefore mainly
motivated as a (fairly crude) way to account for a multiple trials
factor (MTF, Section~\ref{lee}) in searches for phenomena poorly
specified in advance.  However, the threshold had at least one other
likely motivation, namely that in retrospect spurious resonances often
were attributed to mistakes in modeling the detector or other
so-called ``systematic effects'' that were either unknown or not
properly taken into account.  The ``5$\sigma$'' threshold provides
crude protection against such mistakes.

Unfortunately, many current HEP practitioners are unaware of the
original motivation for ``5$\sigma$'', and some may apply this rule
without much thought.  For example, it is sometimes used as a
threshold when an MTF correction (Section~\ref{lee}) has already been
applied, or when there is no MTF from multiple bins or histograms
because the measurement corresponds to a completely specified location
in parameter space, aside from the value of $\theta$ in the composite
hypothesis.  In this case, there is still the question of how many
measurements of other quantities to include in the number of trials
\citep{lyonslee2010}.  Further thoughts on 5$\sigma$ are given in a
recent note by \citet{lyons2013}.

\subsection{Multiple trials factors for scanning nuisance parameters that are
\texorpdfstring{\textit{not}}{not} eliminated}
\label{lee} 

The situation with the MTF described in the previous section can arise
whenever there is nuisance parameter $\psi$ that the analysts choose
not to eliminate, but instead choose to communicate the results
($p$-value and confidence interval for $\theta$) {\em as a function of
  $\psi$}.  The search for the Higgs boson \citep{atlashiggs,cmshiggs}
is such an example, where $\psi$ is the mass of the boson, while
$\theta$ is the Poisson mean (relative to that expected for the SM
Higgs boson) of any putative excess of events at mass $\psi$.  For
each mass $\psi$ there is a $p$-value for the departure from $H_0$,
{\em as if that mass had been fixed in advance}{}, as well as a
confidence interval for $\theta$, given that $\psi$.  This $p$-value
is the ``local'' $p$-value, the probability for a deviation at least
as extreme as that observed, at that {\em particular} mass.  (Local
$p$-values are correlated with those at nearby masses due to
experimental resolution of the mass measurement.)

One can then scan all masses in a specified range and find the
smallest local $p$-value, $\pmin$.  The probability of having a local
$p$-value as small or smaller than $\pmin$, {\em anywhere in a
  specified mass range}, is greater than $\pmin$, by a factor that is
effectively a MTF (also known as the ``Look Elsewhere Effect'' in
HEP).  When feasible, the LHC experiments use Monte Carlo simulations
to calculate the $p$-value that takes this MTF into account, and refer
to that as a ``global'' $p$-value for the specified mass range.  When
this is too computationally demanding, they estimate the global
$p$-value using the method advocated by \citet*{grossvitells2010},
which is based on that of \citet{davies1987}.

To emphasize that the range of masses used for this effective MTF is
arbitrary or subjective, and to indicate the sensitivity to the range,
the LHC collaborations chose to give the global $p$-value for two
ranges of mass (\citet[pp.\ 11,14]{atlashiggs} and \citet[pp.\
33,41]{cmshiggs}).  Some possibilities were the range of masses for
which the SM Higgs boson was not previously ruled out at high
confidence; the range of masses for which the experiment is capable of
observing the SM Higgs boson; or the range of masses for which
sufficient data had been acquired to search for any new boson.  The
collaborations made different choices.

\section{Can results of hypothesis tests be cross-calibrated among different searches?}
\label{pvaluesummary}

In communicating the results of an experiment, generally the goal is
to describe the methods, data analysis, and results, as well as the
authors' interpretations and conclusions, in a manner that enables
readers to draw their own conclusions.  Although at times authors
provide a description of the likelihood function for their
observations, it is common to assume that confidence intervals (often
given for more than one confidence level) and $p$-values (frequently
expressed as equivalent $z$ of Eqn.~\ref{eqn:z}) are sufficient input
into inferences or decisions to be made by readers.

It can therefore be asked what is the result of an author (or reader)
taking the $p$-value as the ``observed data'' for a full (subjective)
Bayesian calculation of the posterior probability of $H_0$.  One could
even attempt to go further and formulate a decision on whether to
claim publicly that $H_0$ is false, using a (subjective) loss function
describing one's personal costs of falsely declaring a discovery,
compared to not declaring a true discovery.

From Eqn.~\ref{scaling}, clearly $z$ alone is not sufficient to
recover the Bayes factor and proceed as a Bayesian.  This point is
repeatedly emphasized in articles already cited.  (Even worse is to
try to recover the BF using only the binary inputs as to whether the
$p$-value was above some fixed thresholds
\citep{dickey1977,bergermortera1991,johnstone1995}.)  The oft-repeated
argument (e.g., \citet[p.~143]{raftery1995}) is that there is no
justification for the step in the derivation of the $p$-value where
``probability density for data as extreme as that observed'' is
replaced with ``probability for data as extreme, {\em or more
  extreme}''.  \citet[p.~385]{jeffreys1961} still seems to be
unsurpassed in his ironic way of saying this (italics in original),
``{\em What the use of [the $p$-value] implies, therefore, is that a
  hypothesis that may be true may be rejected because it has not
  predicted observable results that have not occurred.}''

\citet{good1992} opined that, ``The real objection to [$p$-values] is
not that they usually are utter nonsense, but rather that they can be
highly misleading, especially if the value of [$n$] is not also taken
into account and is large.''  He suggested a rule of thumb for taking
$n$ into account by standardizing the $p$-value to an effective size
of $n=100$, but this seems not to have attracted a following.

Meanwhile, often a confidence interval for $\theta$ (as invariably
reported in HEP publications for 68\% CL and at times for other
values) {\em does} give a good sense of the magnitude of $\sigmatot$
(although this might be misleading in certain special cases).  And one
has a subjective prior and therefore its scale $\tau$.  {\em Thus, at
  least crudely, the required inputs are in hand to recover the result
  from something like Eqn.~\ref{scaling}.}  It is perhaps doubtful
that most physicists would use them to arrive at the same Ockham
factor as calculated through a BF from the original likelihood
function.  On the other hand, a BF based on an arbitrary
(``objective'') $\tau$ does not seem to be an obviously better way to
communicate a result.

While the ``5$\sigma$'' criterion in HEP gets a lot of press, I think
that when a decision needs to be made, physicists intuitively and
informally adjust their decision-making based on the $p$-value, the
confidence interval, their prior belief in $H_0$ and $g(\theta)$, and
their personal sense of costs and risks.

\section{Summary and Conclusions}
\label{conclusion}
More than a half century after Lindley drew attention to the different
dependence of $p$-values and Bayes factors on sample size $n$
(described two decades previously by Jeffreys), there is still no
consensus on how best to communicate results of testing scientific
hypotheses. The argument continues, especially within the broader
Bayesian community, where there is much criticism of $p$-values, and
praise for the ``logical'' approach of Bayes factors.  A core issue
for scientific communication is that the Ockham factor
$\sigmatot/\tau$ is either arbitrary or personal, even asymptotically
for large $n$.

It has always been important in Bayesian point and interval estimation
for the analyst to describe the sensitivity of results to choices of
prior probability, especially for problems involving many parameters.
In testing hypotheses, such sensitivity analysis is clearly mandatory.
The issue is not really the difference in numerical value of
$p$-values and posterior probabilities (or Bayes factors) as one must
commit the error of transposing the conditional probability (fallacy
of probability inversion) to equate the two.  Rather, the fundamental
question is whether a summary of the experimental results, with say
two or three numbers, can (even in principle) be interpreted in a
manner cross-calibrated across different experiments.  The difference
in scaling with sample size (or more generally, the difference in
scaling with $\sigmatot/\tau$) of the \BF and likelihood ratio
$\lambda$ is already apparent in Eqn.~\ref{ockham}; therefore the
additional issue of tail probabilities of data not observed, pithily
derided by Jeffreys (Section~\ref{pvaluesummary} above), cannot bear
all the blame for the paradox.

It is important to gain more experience in HEP with Bayes factors, and
also with Bernardo's intriguing proposals.  For statisticians, I hope
that this discussion of the issues in HEP provides ``existence
proofs'' of situations where we cannot ignore the \JL paradox, and
renews some attempts to improve methods of scientific communication.

\medskip 

{\bf Acknowledgments} I thank my colleagues in high energy physics,
and in the CMS collaboration in particular, for many useful
discussions. I am grateful to members of the CMS Statistics Committee
for comments on an early draft of the manuscript, and in particular to
Luc Demortier and Louis Lyons for continued discussions.  Tom Ferbel
provided invaluable detailed comments on two previous versions that I
posted on the arXiv.  The PhyStat series of workshops organized by
Louis Lyons has led to many fruitful discussions and enlightening
contact with prominent members of the statistics community.  This
material is based upon work partially supported by the U.S. Department
of Energy under Award Number DE-SC0009937.

\bibliographystyle{spbasic}      
\bibliography{JeffLindHigg}

\begin{thebibliography}{107}
\providecommand{\natexlab}[1]{#1}
\providecommand{\url}[1]{{#1}}
\providecommand{\urlprefix}{URL }
\expandafter\ifx\csname urlstyle\endcsname\relax
  \providecommand{\doi}[1]{DOI~\discretionary{}{}{}#1}\else
  \providecommand{\doi}{DOI~\discretionary{}{}{}\begingroup
  \urlstyle{rm}\Url}\fi
\providecommand{\eprint}[2][]{\url{#2}}

\bibitem[{Aad et~al(2012)}]{atlashiggs}
Aad G, et~al (2012) Observation of a new particle in the search for the
  {S}tandard {M}odel {H}iggs boson with the {ATLAS} detector at the {LHC}.
  Physics Letters B 716(1):1--29, \doi{10.1016/j.physletb.2012.08.020}

\bibitem[{Aad et~al(2013)}]{atlashiggsprop2013}
Aad G, et~al (2013) Measurements of {H}iggs boson production and couplings in
  diboson final states with the {ATLAS} detector at the {LHC}. Physics Letters
  B 726:88--119, \doi{10.1016/j.physletb.2013.08.010}

\bibitem[{Aaij et~al(2013)}]{lhcb-bsubs}
Aaij R, et~al (2013) Measurement of the {$B^0_s \to \mu^+ \mu^-$} branching
  fraction and search for {$B^0 \to \mu^+ \mu^-$} decays at the {LHCb}
  experiment. Phys Rev Lett 111:101805, \doi{10.1103/PhysRevLett.111.101805}

\bibitem[{Aaltonen et~al(2009)}]{cdf-singletop}
Aaltonen T, et~al (2009) First observation of electroweak single top quark
  production. Phys Rev Lett 103:092002, \doi{10.1103/PhysRevLett.103.092002}

\bibitem[{Abazov et~al(2009)}]{d0-singletop}
Abazov V, et~al (2009) Observation of single top quark production. Phys Rev
  Lett 103:092,001, \doi{10.1103/PhysRevLett.103.092001}

\bibitem[{Alvarez(1968)}]{alvarez1968}
Alvarez L (1968) Nobel lecture: Recent developments in particle physics.
  \urlprefix\url{http://www.nobelprize.org/nobel_prizes/physics/laureates/1968/alvarez-lecture.html}

\bibitem[{Anderson(1992)}]{anderson1992}
Anderson PW (1992) The {R}everend {T}homas {B}ayes, needles in haystacks, and
  the fifth force. Physics Today 45(1):9--11, \doi{10.1063/1.2809482}

\bibitem[{Andrews(1994)}]{andrews1994}
Andrews DWK (1994) The large sample correspondence between classical hypothesis
  tests and {B}ayesian posterior odds tests. Econometrica 62(5):1207--1232,
  \urlprefix\url{http://www.jstor.org/stable/2951513}

\bibitem[{APA(2010)}]{apa2010}
APA (2010) Publication Manual of the {A}merican {P}sychological {A}ssociation,
  6th edn. American Psychological Association, Washington, DC

\bibitem[{Arisaka et~al(1993)}]{e791-1993}
Arisaka K, et~al (1993) Improved upper limit on the branching ratio
  {$B(K^0_L\rightarrow\mu^\pm e^\mp$}. Phys Rev Lett 70:1049--1052,
  \doi{10.1103/PhysRevLett.70.1049}

\bibitem[{Babu et~al(2013)}]{babu2013}
Babu K, et~al (2013) Baryon number violation, \eprint{arXiv:1311.5285 [hep-ph]}

\bibitem[{Baker and Cousins(1984)}]{bakercousins1984}
Baker S, Cousins RD (1984) Clarification of the use of chi-square and
  likelihood functions in fits to histograms. Nucl Instrum Meth 221:437--442,
  \doi{10.1016/0167-5087(84)90016-4}

\bibitem[{Barroso et~al(1984)Barroso, Branco, and Bento}]{barroso1984}
Barroso A, Branco G, Bento M (1984) $\mathrm{{K}^0_{L}}\rightarrow\bar\mu e$:
  {C}an it be observed? Physics Letters B 134(1-2):123 -- 127,
  \doi{http://dx.doi.org/10.1016/0370-2693(84)90999-7}

\bibitem[{Bartlett(1957)}]{bartlett1957}
Bartlett MS (1957) A comment on {D}. {V}. {L}indley's statistical paradox.
  Biometrika 44(3/4):533--534,
  \urlprefix\url{http://www.jstor.org/stable/2332888}

\bibitem[{Bayarri(1987)}]{bayarri1987}
Bayarri MJ (1987) [{T}esting precise hypotheses]: Comment. Statistical Science
  2(3):342--344, \urlprefix\url{http://www.jstor.org/stable/2245776}

\bibitem[{Bayarri et~al(2012)Bayarri, Berger, Forte, and
  García-Donato}]{bayarri2012}
Bayarri MJ, Berger JO, Forte A, García-Donato G (2012) Criteria for {B}ayesian
  model choice with application to variable selection. The Annals of Statistics
  40(3):1550--1577, \urlprefix\url{http://www.jstor.org/stable/41713685}

\bibitem[{Berger(2008)}]{berger2007phystat}
Berger J (2008) A comparison of testing methodologies. In: Prosper H, Lyons L,
  {De Roeck} A (eds) Proceedings of {PHYSTAT LHC} Workshop on Statistical
  Issues for {LHC} Physics, {CERN}, {G}eneva, {S}witzerland, 27-29 {J}une 2007,
  CERN, CERN-2008-001, pp 8--19,
  \urlprefix\url{http://cds.cern.ch/record/1021125}

\bibitem[{Berger(2011)}]{berger2011phystat}
Berger J (2011) The {B}ayesian approach to discovery. In: Prosper HB, Lyons L
  (eds) Proceedings of {PHYSTAT} 2011 Workshop on Statistical Issues Related to
  Discovery Claims in Search Experiments and Unfolding, {CERN}, {G}eneva,
  {S}witzerland, 17-20 {J}anuary 2011, CERN, CERN-2011-006, pp 17--26,
  \urlprefix\url{http://cdsweb.cern.ch/record/1306523}

\bibitem[{Berger(1985)}]{bergerdecision1985}
Berger JO (1985) Statistical Decision Theory and Bayesian Analysis, 2nd edn.
  Springer Series in Statistics, Springer, New York

\bibitem[{Berger and Delampady(1987{\natexlab{a}})}]{bergerdelampady1987}
Berger JO, Delampady M (1987{\natexlab{a}}) Testing precise hypotheses.
  Statistical Science 2(3):317--335,
  \urlprefix\url{http://www.jstor.org/stable/2245772}

\bibitem[{Berger and Delampady(1987{\natexlab{b}})}]{bergerdelampady1987r}
Berger JO, Delampady M (1987{\natexlab{b}}) [{T}esting precise hypotheses]:
  Rejoinder. Statistical Science 2(3):348--352,
  \urlprefix\url{http://www.jstor.org/stable/2245779}

\bibitem[{Berger and Mortera(1991)}]{bergermortera1991}
Berger JO, Mortera J (1991) Interpreting the stars in precise hypothesis
  testing. International Statistical Review / Revue Internationale de
  Statistique 59(3):337--353,
  \urlprefix\url{http://www.jstor.org/stable/1403691}

\bibitem[{Berger and Pericchi(2001)}]{bergerpericchi2001}
Berger JO, Pericchi LR (2001) Objective {B}ayesian methods for model selection:
  Introduction and comparison. Lecture Notes-Monograph Series 38:135--207,
  \urlprefix\url{http://www.jstor.org/stable/4356165}

\bibitem[{Berger and Sellke(1987)}]{bergersellke1987}
Berger JO, Sellke T (1987) Testing a point null hypothesis: The
  irreconcilability of p values and evidence. Journal of the American
  Statistical Association 82(397):112--122,
  \urlprefix\url{http://www.jstor.org/stable/2289131}

\bibitem[{Berkson(1938)}]{berkson1938}
Berkson J (1938) Some difficulties of interpretation encountered in the
  application of the chi-square test. Journal of the American Statistical
  Association 33(203):526--536,
  \urlprefix\url{http://www.jstor.org/stable/2279690}

\bibitem[{Bernardo(1999)}]{bernardo1999bayes}
Bernardo JM (1999) Nested hypothesis testing: The {B}ayesian reference
  criterion. In: Bernardo JM, Berger JO, Dawid AP, Smith AFM (eds) Bayesian
  Statistics 6. Proceedings of the Sixth Valencia International Meeting, Oxford
  U. Press, Oxford, U.K., pp 101--130

\bibitem[{Bernardo(2009)}]{bernardo2009}
Bernardo JM (2009) [{H}arold {J}effreys's theory of probability revisited]:
  Comment. Statistical Science 24(2):173--175,
  \urlprefix\url{http://www.jstor.org/stable/25681292}

\bibitem[{Bernardo(2011{\natexlab{a}})}]{bernardo2011phystat}
Bernardo JM (2011{\natexlab{a}}) {B}ayes and discovery: Objective {B}ayesian
  hypothesis testing. In: Prosper HB, Lyons L (eds) Proceedings of {PHYSTAT}
  2011 Workshop on Statistical Issues Related to Discovery Claims in Search
  Experiments and Unfolding, {CERN}, {G}eneva, {S}witzerland, 17-20 {J}anuary
  2011, CERN-2011-006, pp 27--49,
  \urlprefix\url{http://cdsweb.cern.ch/record/1306523}

\bibitem[{Bernardo(2011{\natexlab{b}})}]{bernardo2011bayes}
Bernardo JM (2011{\natexlab{b}}) Integrated objective {B}ayesian estimation and
  hypothesis testing. In: Bernardo JM, Bayarri MJ, Berger JO, Dawid AP,
  Heckerman D, Smith AFM, West M (eds) Bayesian Statistics 9. Proceedings of
  the Ninth Valencia International Meeting, Oxford U. Press, Oxford, U.K., pp
  1--68, \urlprefix\url{http://www.uv.es/bernardo/}

\bibitem[{Bernardo and Rueda(2002)}]{bernardorueda2002}
Bernardo JM, Rueda R (2002) {B}ayesian hypothesis testing: A reference
  approach. International Statistical Review / Revue Internationale de
  Statistique 70(3):351--372,
  \urlprefix\url{http://www.jstor.org/stable/1403862}

\bibitem[{Box(1976)}]{box1976}
Box GEP (1976) Science and statistics. Journal of the American Statistical
  Association 71(356):791--799,
  \urlprefix\url{http://www.jstor.org/stable/2286841}

\bibitem[{Box(1980)}]{box1980}
Box GEP (1980) Sampling and {B}ayes' inference in scientific modelling and
  robustness. Journal of the Royal Statistical Society Series A (General)
  143(4):pp. 383--430, \urlprefix\url{http://www.jstor.org/stable/2982063}

\bibitem[{Casella and Berger(1987{\natexlab{a}})}]{casellaberger1987}
Casella G, Berger RL (1987{\natexlab{a}}) Reconciling {B}ayesian and
  frequentist evidence in the one-sided testing problem. Journal of the
  American Statistical Association 82(397):106--111,
  \urlprefix\url{http://www.jstor.org/stable/2289130}

\bibitem[{Casella and Berger(1987{\natexlab{b}})}]{casellaberger1987c}
Casella G, Berger RL (1987{\natexlab{b}}) [{T}esting precise hypotheses]:
  Comment. Statistical Science 2(3):344--347,
  \urlprefix\url{http://www.jstor.org/stable/2245777}

\bibitem[{CERN(2013)}]{PR-bsubs}
CERN (2013) {CERN} experiments put {S}tandard {M}odel to stringent test. \\
  \urlprefix\url{http://press.web.cern.ch/press-releases/2013/07/cern-experiments-put-standard-model-stringent-test}

\bibitem[{Chatrchyan et~al(2012)}]{cmshiggs}
Chatrchyan S, et~al (2012) Observation of a new boson at a mass of 125 {GeV}
  with the {CMS} experiment at the {LHC}. Physics Letters B 716(1):30 -- 61,
  \doi{10.1016/j.physletb.2012.08.021}

\bibitem[{Chatrchyan et~al(2013{\natexlab{a}})}]{cms-bsubs}
Chatrchyan S, et~al (2013{\natexlab{a}}) Measurement of the {$B^0_s \to \mu^+
  \mu^-$} branching fraction and search for {$B^0 \to \mu^+ \mu^-$} with the
  {CMS} experiment. Phys Rev Lett 111:{101804},
  \doi{10.1103/PhysRevLett.111.101804}

\bibitem[{Chatrchyan et~al(2013{\natexlab{b}})}]{cmshiggsprop2012}
Chatrchyan S, et~al (2013{\natexlab{b}}) Study of the mass and spin-parity of
  the {H}iggs boson candidate via its decays to {$Z$} boson pairs. Phys Rev
  Lett 110:081803, \doi{10.1103/PhysRevLett.110.081803}

\bibitem[{Chatrchyan et~al(2014)}]{cmshiggsprop4lep2013}
Chatrchyan S, et~al (2014) {Measurement of the properties of a Higgs boson in
  the four-lepton final state}. Phys Rev D 89:092007,
  \doi{10.1103/PhysRevD.89.092007}

\bibitem[{Cousins(2005)}]{cousinsoxford2005}
Cousins RD (2005) Treatment of nuisance parameters in high energy physics, and
  possible justifications and improvements in the statistics literature. In:
  Lyons L, Unel MK (eds) Proceedings of {PHYSTAT 05} Statistical Problems in
  Particle Physics, Astrophysics and Cosmology, {O}xford, {U.K}, {S}eptember
  12-15, 2005, Imperial College Press, pp 75--85,
  \urlprefix\url{http://www.physics.ox.ac.uk/phystat05/proceedings/}

\bibitem[{Cousins and Highland(1992)}]{cousinshighland1992}
Cousins RD, Highland VL (1992) {Incorporating systematic uncertainties into an
  upper limit}. Nuclear Instruments and Methods A 320:331--335,
  \doi{10.1016/0168-9002(92)90794-5}

\bibitem[{Cowan et~al(2011)Cowan, Cranmer, Gross, and Vitells}]{ccgv2011}
Cowan G, Cranmer K, Gross E, Vitells O (2011) Asymptotic formulae for
  likelihood-based tests of new physics. Eur Phys J C 71:1554,
  \doi{10.1140/epjc/s10052-011-1554-0}

\bibitem[{Cox(2006)}]{cox2006}
Cox DR (2006) Principles of Statistical Inference. Cambridge University Press,
  Cambridge

\bibitem[{Davies(1987)}]{davies1987}
Davies RB (1987) Hypothesis testing when a nuisance parameter is present only
  under the alternative. Biometrika 74(1):33--43

\bibitem[{Demortier(2011)}]{demortier2011phystat}
Demortier L (2011) Open issues in the wake of {B}anff 2010. In: Prosper HB,
  Lyons L (eds) Proceedings of {PHYSTAT} 2011 Workshop on Statistical Issues
  Related to Discovery Claims in Search Experiments and Unfolding, {CERN},
  {G}eneva, {S}witzerland, 17-20 {J}anuary 2011, CERN-2011-006, pp 1--11,
  \urlprefix\url{http://cdsweb.cern.ch/record/1306523}

\bibitem[{Dickey(1977)}]{dickey1977}
Dickey JM (1977) Is the tail area useful as an approximate {B}ayes factor?
  Journal of the American Statistical Association 72(357):138--142,
  \doi{10.1080/01621459.1977.10479922},
  \urlprefix\url{http://www.jstor.org/stable/2286921}

\bibitem[{Eadie et~al(1971)}]{eadie1971}
Eadie W, et~al (1971) Statistical Methods in Experimental Physics, 1st edn.
  North Holland, Amsterdam

\bibitem[{Edwards et~al(1963)Edwards, Lindman, and Savage}]{edwards1963}
Edwards W, Lindman H, Savage LJ (1963) {B}ayesian statistical inference for
  psychological research. Psychological Review 70(3):193--242

\bibitem[{Feldman and Cousins(1998)}]{feldman1998}
Feldman GJ, Cousins RD (1998) Unified approach to the classical statistical
  analysis of small signals. Phys Rev D 57:3873--3889,
  \doi{10.1103/PhysRevD.57.3873}, \eprint{physics/9711021}

\bibitem[{Ferguson and Heene(2012)}]{ferguson2012}
Ferguson CJ, Heene M (2012) A vast graveyard of undead theories: Publication
  bias and psychological science's aversion to the null. Perspectives on
  Psychological Science 7(6):555--561, \doi{10.1177/1745691612459059}

\bibitem[{Fermilab(2009)}]{PR-singletop}
Fermilab (2009) Fermilab collider experiments discover rare single top quark.
  \urlprefix
  \url{http://www.fnal.gov/pub/presspass/press_releases/Single-Top-Quark-March2009.html}

\bibitem[{Galison(1983)}]{galison1983}
Galison P (1983) How the first neutral-current experiments ended. Rev Mod Phys
  55:477--509, \doi{10.1103/RevModPhys.55.477}

\bibitem[{Gelman and Rubin(1995)}]{gelmanrubin1995}
Gelman A, Rubin DB (1995) Avoiding model selection in {B}ayesian social
  research. Sociological Methodology 25:165--173,
  \urlprefix\url{http://www.jstor.org/stable/271064}

\bibitem[{Georgi(1993)}]{georgi1993}
Georgi H (1993) {Effective field theory}. Ann Rev Nucl Part Sci 43:209--252,
  \doi{10.1146/annurev.ns.43.120193.001233}

\bibitem[{Good(1992)}]{good1992}
Good IJ (1992) The {B}ayes/non-{B}ayes compromise: A brief review. Journal of
  the American Statistical Association 87(419):597--606,
  \urlprefix\url{http://www.jstor.org/stable/2290192}

\bibitem[{Gross and Vitells(2010)}]{grossvitells2010}
Gross E, Vitells O (2010) Trial factors or the look elsewhere effect in high
  energy physics. Eur Phys J C 70:525--530,
  \doi{10.1140/epjc/s10052-010-1470-8}

\bibitem[{Hasert et~al(1973)}]{hasert1973}
Hasert F, et~al (1973) Observation of neutrino-like interactions without muon
  or electron in the {G}argamelle neutrino experiment. Physics Letters B
  46(1):138 -- 140, \doi{10.1016/0370-2693(73)90499-1}

\bibitem[{Hirsch et~al(2013)Hirsch, P\"as, and Porod}]{sciam2013}
Hirsch M, P\"as H, Porod W (2013) Ghostly beacons of new physics. Scientific
  American 308(April):40--47, \doi{10.1038/scientificamerican0413-40}

\bibitem[{Incandela and Gianotti(2012)}]{july4}
Incandela J, Gianotti F (2012) Latest update in the search for the {H}iggs
  boson, public seminar at {CERN}. Video:
  \url{http://cds.cern.ch/record/1459565}; slides:
  \url{http://indico.cern.ch/conferenceDisplay.py?confId=197461}

\bibitem[{James(1980)}]{james1980}
James F (1980) Interpretation of the shape of the likelihood function around
  its minimum. Comput Phys Commun 20:29--35, \doi{10.1016/0010-4655(80)90103-4}

\bibitem[{James(2006)}]{james2006}
James F (2006) Statistical Methods in Experimental Physics, 2nd edn. World
  Scientific, Singapore

\bibitem[{Jaynes(2003)}]{jaynes2003}
Jaynes E (2003) Probability Theory: The Logic of Science. Cambridge University
  Press, Cambridge, U.K.

\bibitem[{Jeffreys(1961)}]{jeffreys1961}
Jeffreys H (1961) Theory of Probability, 3rd edn. Oxford University Press,
  Oxford

\bibitem[{Johnstone and Lindley(1995)}]{johnstone1995}
Johnstone D, Lindley D (1995) {B}ayesian inference given data `significant at
  $\alpha$': Tests of point hypotheses. Theory and Decision 38(1):51--60,
  \doi{10.1007/BF01083168}

\bibitem[{Kadane(1987)}]{kadane1987}
Kadane JB (1987) [{T}esting precise hypotheses]: Comment. Statistical Science
  2(3):347--348, \urlprefix\url{http://www.jstor.org/stable/2245778}

\bibitem[{Kass(2009)}]{kass2009}
Kass R (2009) Comment: The importance of {J}effreys's legacy. Statistical
  Science 24(2):179--182, \urlprefix\url{http://www.jstor.org/stable/25681294}

\bibitem[{Kass and Raftery(1995)}]{kassraftery1995}
Kass RE, Raftery AE (1995) Bayes factors. Journal of the American Statistical
  Association 90(430):773--795,
  \urlprefix\url{http://www.jstor.org/stable/2291091}

\bibitem[{Kass and Wasserman(1995)}]{kasswasserman1995}
Kass RE, Wasserman L (1995) A reference {B}ayesian test for nested hypotheses
  and its relationship to the {S}chwarz criterion. Journal of the American
  Statistical Association 90(431):928--934,
  \urlprefix\url{http://www.jstor.org/stable/2291327}

\bibitem[{Kirk(1996)}]{kirk1996}
Kirk RE (1996) Practical significance: A concept whose time has come.
  Educational and Psychological Measurement 56(5):746--759,
  \doi{10.1177/0013164496056005002}

\bibitem[{Leamer(1978)}]{leamer1978}
Leamer EE (1978) Specification Searches: Ad Hoc Inference with Nonexperimental
  Data. Wiley series in probability and mathematical statistics, Wiley, New
  York

\bibitem[{Lee(2004)}]{lee2004}
Lee PM (2004) Bayesian Statistics: An Introduction, 3rd edn. Wiley, Chichester
  U.K.

\bibitem[{Lehmann and Romero(2005)}]{lehmann2005}
Lehmann E, Romero JP (2005) Testing Statistical Hypotheses, 3rd edn. Springer,
  New York

\bibitem[{Lindley(2009)}]{lindley2009}
Lindley D (2009) [{H}arold {J}effreys's theory of probability revisited]:
  Comment. Statistical Science 24(2):183--184,
  \urlprefix\url{http://www.jstor.org/stable/25681295}

\bibitem[{Lindley(1957)}]{lindley1957}
Lindley DV (1957) A statistical paradox. Biometrika 44(1/2):187--192,
  \urlprefix\url{http://www.jstor.org/stable/2333251}

\bibitem[{Lyons(2010)}]{lyonslee2010}
Lyons L (2010) Comments on `look elsewhere effect'. \\
  \url{http://www.physics.ox.ac.uk/Users/lyons/LEE_feb7_2010.pdf}

\bibitem[{Lyons(2013)}]{lyons2013}
Lyons L (2013) Discovering the significance of 5 sigma,
  \eprint{arXiv:1310.1284}

\bibitem[{Mayo and Spanos(2006)}]{mayospanos2006}
Mayo DG, Spanos A (2006) Severe testing as a basic concept in a
  {N}eyman-{P}earson philosophy of induction. The British Journal for the
  Philosophy of Science 57(2):pp. 323--357,
  \urlprefix\url{http://www.jstor.org/stable/3873470}

\bibitem[{Nakagawa and Cuthill(2007)}]{nakagawa2007}
Nakagawa S, Cuthill IC (2007) Effect size, confidence interval and statistical
  significance: a practical guide for biologists. Biological Reviews
  82(4):591--605, \doi{10.1111/j.1469-185X.2007.00027.x}

\bibitem[{Neyman(1937)}]{neyman1937}
Neyman J (1937) Outline of a theory of statistical estimation based on the
  classical theory of probability. Philosophical Transactions of the Royal
  Society of London Series A, Mathematical and Physical Sciences 236(767):pp.
  333--380, \urlprefix\url{http://www.jstor.org/stable/91337}

\bibitem[{Neyman and Pearson(1933{\natexlab{a}})}]{neymanpearson1933a}
Neyman J, Pearson ES (1933{\natexlab{a}}) On the problem of the most efficient
  tests of statistical hypotheses. Philosophical Transactions of the Royal
  Society of London Series A, Containing Papers of a Mathematical or Physical
  Character 231:289--337, \urlprefix\url{http://www.jstor.org/stable/91247}

\bibitem[{Neyman and Pearson(1933{\natexlab{b}})}]{neymanpearson1933b}
Neyman J, Pearson ES (1933{\natexlab{b}}) The testing of statistical hypotheses
  in relation to probabilities a priori. Mathematical Proceedings of the
  Cambridge Philosophical Society 29:492--510, \doi{10.1017/S030500410001152X}

\bibitem[{Philippe and Robert(1998)}]{philippe1998}
Philippe A, Robert C (1998) A note on the confidence properties of reference
  priors for the calibration model. Sociedad de Estad\'istica e Investigaci\'on
  Operativa Test 7(1):147--160, \doi{10.1007/BF02565107}

\bibitem[{Prescott et~al(1978)}]{prescott1978}
Prescott C, et~al (1978) Parity non-conservation in inelastic electron
  scattering. Physics Letters B 77(3):347 -- 352,
  \doi{10.1016/0370-2693(78)90722-0}

\bibitem[{Raftery(1995{\natexlab{a}})}]{raftery1995}
Raftery AE (1995{\natexlab{a}}) Bayesian model selection in social research.
  Sociological Methodology 25:111--163,
  \urlprefix\url{http://www.jstor.org/stable/271063}

\bibitem[{Raftery(1995{\natexlab{b}})}]{raftery1995r}
Raftery AE (1995{\natexlab{b}}) Rejoinder: Model selection is unavoidable in
  social research. Sociological Methodology 25:185--195,
  \urlprefix\url{http://www.jstor.org/stable/271066}

\bibitem[{Rice(2007)}]{rice2007}
Rice JA (2007) Mathematical Statistics and Data Analysis, 3rd edn. Thomson,
  Belmont, CA

\bibitem[{Robert(1993)}]{robert1993}
Robert CP (1993) A note on {J}effreys-{L}indley paradox. Statistica Sinica
  3(2):601--608

\bibitem[{Robert(2013)}]{robert2013}
Robert CP (2013) On the {J}effreys-{L}indley paradox, \eprint{arXiv:1303.5973v3
  [stat.ME]}

\bibitem[{Robert et~al(2009)Robert, Chopin, and Rousseau}]{robert2009}
Robert CP, Chopin N, Rousseau J (2009) Harold {J}effreys's theory of
  probability revisited. Statistical Science 24(2):141--172,
  \urlprefix\url{http://www.jstor.org/stable/25681291}

\bibitem[{Rosenfeld(1968)}]{rosenfeld1968}
Rosenfeld AH (1968) Are there any far-out mesons or baryons? In: Baltay C,
  Rosenfeld AH (eds) Meson spectroscopy: A collection of articles, W.A.
  Benjamin, New York, pp 455--483, {F}rom the preface: based on reviews
  presented at the Conference on Meson Spectroscopy, April 26-27, 1968,
  Philadelphia, PA USA. ``...not, however, intended to be the proceedings...''

\bibitem[{Senn(2001)}]{senn2001}
Senn S (2001) Two cheers for p-values? Journal of Epidemiology and
  Biostatistics 6(2):193--204

\bibitem[{Shafer(1982)}]{shafer1982}
Shafer G (1982) {L}indley's paradox. Journal of the American Statistical
  Association 77(378):325--334,
  \urlprefix\url{http://www.jstor.org/stable/2287244}

\bibitem[{Smith and Spiegelhalter(1980)}]{smithspiegelhalter1980}
Smith AFM, Spiegelhalter DJ (1980) Bayes factors and choice criteria for linear
  models. Journal of the Royal Statistical Society Series B (Methodological)
  42(2):213--220, \urlprefix\url{http://www.jstor.org/stable/2984964}

\bibitem[{Spanos(2013)}]{spanos2013}
Spanos A (2013) Who should be afraid of the {J}effreys-{L}indley paradox?
  Philosophy of Science 80(1):73--93,
  \urlprefix\url{http://www.jstor.org/stable/10.1086/668875}

\bibitem[{Stuart et~al(1999)Stuart, Ord, and Arnold}]{kendall1999}
Stuart A, Ord K, Arnold S (1999) Kendall's Advanced Theory of Statistics,
  vol~2A, 6th edn. Arnold, London, and earlier editions by Kendall and Stuart

\bibitem[{Swedish Academy(2013)}]{nobel2013}
Swedish Academy (2013) Advanced information: {S}cientific background: {T}he
  {BEH}-mechanism, interactions with short range forces and scalar particles.
  \urlprefix\url{http://www.nobelprize.org/nobel_prizes/physics/laureates/2013/advanced.html}

\bibitem[{{'t Hooft}(1976)}]{thooft1976}
{'t Hooft} G (1976) Symmetry breaking through {B}ell-{J}ackiw anomalies. Phys
  Rev Lett 37:8--11, \doi{10.1103/PhysRevLett.37.8}

\bibitem[{{'t Hooft}(1999)}]{thooft1999}
{'t Hooft} G (1999) Nobel lecture: A confrontation with infinity.
  \urlprefix\url{http://www.nobelprize.org/nobel_prizes/physics/laureates/1999/thooft-lecture.html}, {T}his web page has video, slides, and pdf writeup.

\bibitem[{Thompson(2007)}]{thompson2007}
Thompson B (2007) The Nature of Statistical Evidence. Lecture Notes in
  Statistics, Springer, New York

\bibitem[{{van Dyk}(2014)}]{vandyk2014}
{van Dyk} DA (2014) The role of statistics in the discovery of a {H}iggs boson.
  Annual Review of Statistics and Its Application 1(1):41--59,
  \doi{10.1146/annurev-statistics-062713-085841}

\bibitem[{Vardeman(1987)}]{vardeman1987}
Vardeman SB (1987) [{T}esting a point null hypothesis: The irreconcilability of
  p values and evidence]: Comment. Journal of the American Statistical
  Association 82(397):130--131,
  \urlprefix\url{http://www.jstor.org/stable/2289136}

\bibitem[{Webster(1969)}]{webster7}
Webster (1969) {Webster's Seventh New Collegiate Dictionary, based on Webster's
  Third New International Dictionary}. G. and C. Merriam Company, Springfield,
  MA

\bibitem[{Welch and Peers(1963)}]{welchpeers1963}
Welch BL, Peers HW (1963) On formulae for confidence points based on integrals
  of weighted likelihoods. Journal of the Royal Statistical Society Series B
  (Methodological) 25(2):pp. 318--329,
  \urlprefix\url{http://www.jstor.org/stable/2984298}

\bibitem[{Wilczek(2004)}]{wilczek2004}
Wilczek F (2004) Nobel lecture: Asymptotic freedom: From paradox to paradigm.
  \urlprefix\url{http://www.nobelprize.org/nobel_prizes/physics/laureates/2004/wilczek-lecture.html}, {T}his web page has video, slides, and pdf writeup.

\bibitem[{Wilkinson et~al(1999)}]{wilkinson1999}
Wilkinson L, et~al (1999) Statistical methods in psychology journals -
  guidelines and explanations. American Psychologist 54(8):594--604,
  \doi{10.1037//0003-066X.54.8.594}

\bibitem[{Zellner(2009)}]{zellner2009}
Zellner A (2009) [{H}arold {J}effreys's theory of probability revisited]:
  Comment. Statistical Science 24(2):187--190,
  \urlprefix\url{http://www.jstor.org/stable/25681297}

\bibitem[{Zellner and Siow(1980)}]{zellnersiow1980}
Zellner A, Siow A (1980) Posterior odds ratios for selected regression
  hypotheses. Trabajos de Estadistica Y de Investigacion Operativa
  31(1):585--603, \doi{10.1007/BF02888369}

\end{thebibliography}

\end{document}